\documentstyle[12pt]{article}
\begin{document}

\hspace*{63ex} OCU-PHYS-158

\hspace*{63ex} October 1995

\begin{center}
{\Large {\bf Energy and pressure densities of a hot quark-gluon
plasma }} \\

\hspace*{3ex}

\hspace*{3ex}

\hspace*{3ex}

{\large {\sc Y. Hisamatsu\footnote{
Present address: Ehime Prefectural Science
Museum, Niihama, Ehime 792, Japan.} and A. Ni\'{e}gawa}\footnote{
E-mail: h1143@ocugw.cc.osaka-cu.ac.jp}

{\normalsize\em Department of Physics, Osaka City University } \\
{\normalsize\em Sumiyoshi-ku, Osaka 558, Japan} } \\

\hspace*{2ex}

\hspace*{2ex}

\hspace*{2ex}

{\large {\bf Abstract}} \\
\end{center}

We calculate the energy and hydrostatic pressure densities of a hot
quark-gluon plasma in thermal equilibrium through diagrammatic
analyses of the statistical average, $\langle \Theta_{\mu \nu}
\rangle$, of the energy-momentum-tensor operator $\Theta_{\mu \nu}$.
To leading order at high temperature, the energy density of the long
wave length modes is consistently extracted by applying the
hard-thermal-loop resummation scheme to the operator-inserted no-leg
thermal amplitudes $\langle \Theta_{\mu \nu} \rangle$. We find that,
for the long wave length gluons, the energy density, being positive,
is tremendously enhanced as compared to the noninteracting case,
while, for the quarks, no noticeable deviation from the
noninteracting case is found.

\newpage
\section{Introduction}
For the long wave length \{ $\lambda$ $\leq$ ${\cal O}
((g T)^{- 1}$) \} or soft modes in a hot quark-gluon plasma, the
contributions of \lq\lq hard thermal loops'' to their amplitudes are
as important as tree amplitudes \cite{pis1,bra1,bra2,fre1}. Numerous
applications have been made \cite{app} of the hard thermal loop
(HTL) resummation scheme \cite{pis1,bra1,bra2,fre1}, the scheme
which enables consistent evaluations of any thermal amplitudes to
leading order in the coupling constant $g$.

Due to the HTL resummations, the energy and pressure densities of
the soft modes in the quark-gluon plasma deviates from the form in
the noninteracting case. Weldon \cite{wel0} was the first who treat
this issue. Subsequently, several work \cite{fre-bla,nai,blaizot}
have been devoted to the issue. The strategy for extracting the
energy-momentum-tensor operator for soft modes in
\cite{wel0,fre-bla,blaizot} and the energy functional in \cite{nai}
is to use the effective action \cite{eff-act1,eff-act2}, in which
all the contributions coming from HTL resummations are taken in. A
brief account is given in \cite{blaizot} of the energy-momentum
tensors derived in \cite{wel0,fre-bla,nai}.

The purpose of the present paper is to evaluate the energy density
${\cal E} (p)$ and the pressure density ${\cal P} (p)$ of the soft
modes in the hot quark-gluon plasma from purely \lq\lq diagrammatic
point of view''. By \lq\lq the soft mode'' we mean the mode whose
momentum $p$ is in the range
\begin{equation}
g^2 T << p << T \, .
\label{soft-region}
\end{equation}
For analyzing the hyper-soft modes $p \leq {\cal O} (g^2 T)$, a
treatment beyond the HTL resummation scheme is required. Our
procedure is as follows: 1) We start with the energy-momentum-tensor
operator, $\Theta_{\mu \nu}$, which are extracted from the QCD
Lagrangian. 2) We evaluate the energy and pressure densities
through calculating the statistical averages $\langle \Theta_{0 0}
\rangle$ and $\langle \Theta_{1 1} \rangle$, respectively. 3) In
evaluating these densities for soft modes, to leading order at high
temperature, we apply the HTL resummation scheme to the operator
inserted no-leg thermal amplitudes, $\langle \Theta_{0 0} \rangle$
and $\langle \Theta_{1 1} \rangle$. Through the HTL resummations, in
addition to (formally) the lowest-order diagrams, formally ${\cal O}
(g^2)$ diagrams yield dominant contributions.

It is worth mentioning that, in the approach
\cite{wel0,fre-bla,nai,blaizot}, the HTL resummations are performed
first to construct the effective action and then the
energy-momentum-tensor operator (for soft modes) is extracted. On
the contrary, our approach starts with the energy-momentum-tensor
operator $\Theta_{\mu \nu}$ extracted from the Lagrangian and then
$\langle \Theta_{0 0} \rangle$ and $\langle \Theta_{1 1} \rangle$
are evaluated on the basis of the HTL resummation scheme. It turns
out that the resultant $\langle \Theta_{\mu \nu} \rangle$
corresponds to the thermal average of the energy-momentum-tensor
operator $T_{\mu \nu}$ found in \cite{fre-bla}. Relationships
between various expressions for $T_{\mu \nu}$, found in
\cite{wel0,fre-bla,nai,blaizot}, are not immediately obvious
\cite{blaizot}. This is also the case for the relations of our
$\langle \Theta_{\mu \nu} \rangle$ to  $T_{\mu \nu}$'s in
\cite{wel0,nai,blaizot}.

The plan of the paper is as follows. In Sec. II, some preliminaries,
which include the expression for $\Theta_{\mu \nu}$ extracted from
the Lagrangian, are given. The contribution to $\langle
\Theta_{\mu \nu} \rangle$ from formally lowest-order diagrams are
evaluated in Sec. III. In Sec. IV, on the basis of the
HTL-resummation scheme, we calculate the contributions to $\langle
\Theta_{\mu \nu} \rangle$ from formally ${\cal O} (g^2)$ diagrams.
In Sec. V, the energy of the soft modes is calculated and, in Sec.
VI, the hydrostatic pressure is calculated. Sec. VII is devoted
to summary and discussion.
\section{Preliminary}

We consider an $S U (N)$ gauge theory with $N_f$ flavors of massless
quarks in the fundamental representation. As is well known (see,
e.g., \cite{lan}), various forms of the energy-momentum-tensor
operator are available. Although they lead to different
stress-energy-momentum densities, the same total or integrated
stress energy momentum results. Since we are interested in the
stress-energy-momentum density of the quark-gluon plasma, it is of
vital importance which form of the energy-momentum tensor is the
physically sensible one. A natural candidate is the gravitational
energy-momentum tensor $\Theta_{\mu \nu}$, since its generalization
to the case of curved space-time appears in the Einstein equation.
As in all other known cases, $\Theta_{\mu \nu}$ coincides \cite{lan}
with the Belinfante tensor:
\begin{eqnarray}
\Theta_{\mu \nu} & \equiv & \Theta_{\mu \nu}^{(g)} +
\Theta_{\mu \nu}^{(q)} \, ,
\label{theta} \\
\Theta_{\mu \nu}^{(g)} & = & - F^a_{\mu \rho}
F^{a \rho}_\nu + \frac{1}{4} \, g_{\mu \nu} \,
F^{a \alpha \beta} F^a_{\alpha \beta} \nonumber \\
& & + \left\{ - (\partial_\mu \, \bar{\omega}^a) (D_\nu \, \omega)^a
+ \lambda A_\mu^a \partial_\nu \partial \cdot A^a \right\} + \left\{
\mu \leftrightarrow \nu \right\} \nonumber \\
& & + g_{\mu \nu} \Big[ (\partial^\rho \bar{\omega}^a)
(D_\rho \omega)^a \nonumber \\
& & \mbox{\hspace*{6.2ex}} - \frac{\lambda}{2} (\partial
\cdot A^a)^2 - \lambda A_\rho^a \partial^\rho \partial \cdot A^a
\Big] \, , \label{theta-g} \\
\Theta_{\mu \nu}^{(q)} & = & \frac{i}{2} \bar{\psi} \left(
\gamma_\mu \stackrel{\leftrightarrow}{D}_\nu + \gamma_\nu
\stackrel{\leftrightarrow}{D}_\mu \right) \psi
- i g_{\mu \nu} \bar{\psi} \stackrel{\leftrightarrow}
{{D\kern-0.14em\raise0.17ex\llap{/}\kern0.15em\relax}} \,
\psi \, ,
\label{theta-q}
\end{eqnarray}
where
\begin{eqnarray*}
& & (D_\mu \, \omega)^a \equiv \partial_\mu \omega^a + i g f^{a b c}
A_\mu^b \omega^c \, , \\
& & \stackrel{\leftrightarrow}{D}_\mu \equiv \frac{1}{2}
\stackrel{\leftrightarrow}{\partial}_\mu - g
{A\kern-0.em\raise0.17ex\llap{/}\kern0.15em\relax}^a T^a \, .
\end{eqnarray*}
Here $\{ f^{a b c} \}$ are the structure constants of $s u (N)$ and
$\{ T^a \}$ are a anti-Hermitian basis of the fundamental
representation. The temperature-dependent part of the
energy-momentum ${\bf P}_\mu$ of a quark-gluon plasma is obtained
through
\begin{equation}
\langle \Theta_{\mu \nu} \rangle \equiv \frac{Tr \left( e^{-
H / T} \Theta_{\mu \nu} \right)}{Tr \left( e^{- H / T} \right)}
\label{thermal-ave}
\end{equation}
as
\begin{equation}
{\bf P}_\mu \equiv \langle \Theta_{0 \mu} \rangle - \langle 0
|\Theta_{0 \mu} | 0 \rangle \, ,
\label{P-mu}
\end{equation}
where
\[
\langle 0 | \Theta_{\mu \nu} | 0 \rangle = \lim_{T \to 0}
\langle \Theta_{\mu \nu} \rangle \, .
\]
It is to be noted that, in (\ref{P-mu}), the ultraviolet (UV)
divergence involved in $\langle \Theta_{0 \mu} \rangle$ and that in
$\langle 0 | \Theta_{0 \mu} | 0 \rangle$ cancel out. Similarly the
temperature-dependent part of the hydrostatic pressure ${\bf P}$ is
defined as
\begin{equation}
{\bf P} \equiv \langle \Theta_{1 1} \rangle - \langle 0
| \Theta_{1 1} | 0 \rangle \, .
\label{Pressure}
\end{equation}

We employ the imaginary-time formalism \cite{lan} of thermal field
theory. Nevertheless, all the formulae are displayed in the
Minkowski metric. Following \cite{bra1}, we introduce an index
\lq\lq $r$'', $r = +$ for bosons and $r = -$ for fermions:
\begin{equation}
\frac{1}{P^2} \equiv \Delta^r (P) = \left\{
\begin{array}{ll}
\Delta^+ (P) \, , \;\;\;\;\; & \mbox{for bosons}
\, , \\
\Delta^- (P) \, , \;\;\;\;\; & \mbox{for fermions} \, .
\end{array}
\right.
\label{mode}
\end{equation}
Here, for $r = + \; \{-\}$, $p_0$ takes the values $p_0 = 2 n i \pi
T$ $\{(2 n + 1) i \pi T \}$ with $n = 0, \pm 1, \pm 2, ...\;$.
Capital letters represent four-momenta, lower-case letters their
components: $P_\mu = (p_0, {\bf p})$ with ${\bf p} = p
\hat{{\bf p}}$, and $P^2 = p_0^2 - {\bf p}^2$. For a loop momentum
$P$, we introduce the symbol Tr
\raisebox{-2.6mm}{\scriptsize{$\!\!\!\!\!\!\!\!\!\!\:$} P} defined
as
\begin{equation}
\mbox{Tr} \raisebox{-2.6mm}{\scriptsize{$\!\!\!\!\!\!\!$} P} \;\;
F (p_0, p) \equiv T \sum_{n = - \infty}^{+ \infty} \int
\frac{d^{\, 3} p}{(2 \pi)^3} F (p_0, p) \, .
\label{symbol}
\end{equation}
For bosonic \{fermionic\} $P$, $p_0 = 2 n i \pi T$ $\{(2 n + 1) i
\pi T \}$. $\,$ Tr
$\raisebox{-2.6mm}{\scriptsize{$\!\!\!\!\!\!\!\!\!\!\!\!\!$soft}
P}$ is defined as in (\ref{symbol}), where the integration over
${\bf p}$ is carried out over the soft-$p$ region, Eq.
(\ref{soft-region}).

In this paper, we use Feynman gauge ($\lambda = 1$) throughout.
\section{Contributions from tadpole diagrams to $\langle
\Theta_{\mu \nu} \rangle$}

To the lowest order, (\ref{thermal-ave}) is diagrammed in Fig. 1:
Figs. 1(a), (b), and (c) represent the contributions from soft
gluons, FP-ghosts, and quarks, respectively. Incidentally, for hard
gluons [quarks], the HTL-resummed effective gluon [quark] propagator
--- the line with a blob --- in Fig. 1(a) [(c)] can be replaced by
the bare counterpart. The form of their contributions are well
known, and we do not reproduce them here.
\subsection{Gluon sector}

We start with the gluon sector, Figs. 1(a) and (b). Using
(\ref{theta-g}), we obtain for the contribution to $\langle
\Theta_{\mu \nu} \rangle$,
\begin{eqnarray}
& & \langle \Theta^{(g 0)}_{\mu \nu} \rangle =
\langle \Theta^{(g 0)}_{\mu \nu} \rangle
\rule[-3mm]{.14mm}{7mm} \raisebox{-2mm}{\scriptsize{$\;$Fig. 1(a)}}
+ \langle \Theta^{(g 0)}_{\mu \nu} \rangle
\rule[-3mm]{.14mm}{7mm} \raisebox{-2mm}{\scriptsize{$\;$Fig. 1(b)}}
\, ,
\label{g0} \\
& & \langle \Theta^{(g 0)}_{\mu \nu} \rangle
\rule[-3mm]{.14mm}{7mm} \raisebox{-2mm}{\scriptsize{$\;$Fig. 1(a)}}
= \frac{1}{2} \left( N ^2 - 1 \right) \; \mbox{Tr}
\raisebox{-2.6mm}{\scriptsize{$\!\!\!\!\!\!\!\!\!\!$soft} P}
\displaystyle{\raisebox{0.6ex}{\scriptsize{*}}} \!
\Delta_{\alpha \beta} (P) \, \theta^{\alpha \beta}_{\mu \nu} (P)
\, , \nonumber \\
\label{g0-1} \\
& & \langle \Theta^{(g 0)}_{\mu \nu} \rangle
\rule[-3mm]{.14mm}{7mm} \raisebox{-2mm}{\scriptsize{$\;$Fig. 1(b)}}
\nonumber \\
& & \mbox{\hspace*{4ex}} = - \left( N^2 - 1 \right) \; \mbox{Tr}
\raisebox{-2.6mm}{\scriptsize{$\!\!\!\!\!\!\!\!\!\!$soft} P}
\Delta^+ (P) \left[ g_{\mu \nu} P^2 - 2 P_\mu P_\nu \right] \, .
\label{g0-2}
\end{eqnarray}

\noindent where $\displaystyle{\raisebox{0.6ex}{\scriptsize{*}}} \!
\Delta_{\alpha \beta} (P)$ is the effective gluon propagator in
Feynman gauge,
\begin{equation}
\displaystyle{\raisebox{0.6ex}{\scriptsize{*}}} \!
\Delta_{\alpha \beta} (P) = \frac{P_\alpha P_\beta}{P^4} +
\sum_{\ell = T, \, L} \frac{{\cal Q}^{(\ell)}_{\alpha \beta}}{P^2 -
\Pi_\ell (P)} \, ,
\label{g-eff-pro}
\end{equation}
and, in (\ref{g0-1}),
\begin{eqnarray}
\theta^{\alpha \beta}_{\mu \nu} (P) & = & \left\{ - P_\mu P_\nu
g^{\alpha \beta} - P^2 \, \delta^\alpha_\mu \, \delta^\beta_\nu +
\frac{1}{2} \, g_{\mu \nu} P^2 g^{\alpha \beta} \right\} \nonumber
\\
& & + \left\{ \alpha \leftrightarrow \beta
\right\} \, .
\label{2-pt-theta}
\end{eqnarray}
In (\ref{g-eff-pro}),
\begin{eqnarray}
& & {\cal Q}^{(T)}_{\alpha \beta} = \sum_{i, \, j = 1}^3
g_{\alpha i} \, g_{\beta j} \left( \hat{{\bf p}}_i \,
\hat{{\bf p}}_j - \delta_{i j} \right) \, , \;\;\;\;\;\;\;\;\;\;
(\hat{{\bf p}} \equiv {\bf p}/ |{\bf p}|) \nonumber \, ,
\\
& & {\cal Q}^{(L)}_{\alpha \beta} = g_{\alpha 0} \, g_{\beta 0} -
\frac{P_\alpha P_\beta}{P^2} - \sum_{i, \, j = 1}^3 g_{\alpha i} \,
g_{\beta j} \hat{{\bf p}}_i \, \hat{{\bf p}}_j \, ,
\label{lon-vac}
\end{eqnarray}
and $\Pi_T$ ($\Pi_L$) is the the transverse (longitudinal) part of
the thermal vacuum polarization tensor \cite{gluon}, whose explicit
forms are not necessary for the purpose of this paper.
\subsection{Quark sector}

Fig. 1 (c) with (\ref{theta-q}) gives
\begin{eqnarray}
\langle \Theta_{\mu \nu}^{(q 0)} \rangle & = & \frac{1}{2} N_f \; \,
\mbox{Tr} \raisebox{-2.6mm}{\scriptsize{$\!\!\!\!\!\!\!\!\!\!$soft}
P} \Big[ \left\{ P_\nu \, tr \left[ \gamma_\mu
\displaystyle{\raisebox{0.6ex}{\scriptsize{*}}} \!
{S\kern-0.07em\raise0.17ex\llap{/}\kern0.15em\relax} (P) \right]
\right. \nonumber \\
& & \left. \mbox{\hspace{11.9ex}} - g_{\mu \nu} \, tr \left[
{P\kern-0.07em\raise0.17ex\llap{/}\kern0.15em\relax} \,
\displaystyle{\raisebox{0.6ex}{\scriptsize{*}}} \!
{S\kern-0.07em\raise0.17ex\llap{/}\kern0.15em\relax} (P) \right]
\right\} + \left\{ \mu \leftrightarrow \nu \right\} \Big] \, ,
\nonumber \\
\label{q01}
\end{eqnarray}
\noindent where $\displaystyle{\raisebox{0.6ex}{\scriptsize{*}}} \!
S (P)$ is the effective quark propagator \cite{quark}:
\begin{eqnarray}
\displaystyle{\raisebox{0.6ex}{\scriptsize{*}}} \! S (P) & = &
\sum_{\sigma = \pm}
\hat{P\kern-0.12em\raise0.17ex\llap{/}\kern0.15em\relax}_\sigma
\displaystyle{\raisebox{0.6ex}{\scriptsize{*}}} \!
\tilde{S}^{(\sigma)} (P) \, ,
\label{S-hat} \\
\hat{P}_\sigma & = & (1, \sigma \hat{{\bf p}}) \nonumber \, .
\end{eqnarray}
The explicit form of
$\displaystyle{\raisebox{0.6ex}{\scriptsize{*}}} \!
\tilde{S}^{(\sigma)} (P)$ in (\ref{S-hat}) is given in \cite{quark}.
\section{Hard-thermal-loop resummation for diagrams with operator
$\Theta_{\mu \nu}$ insertion}
\subsection{General observations}

Now we analyze formally ${\cal O} (g^2)$ contributions to $\langle
\Theta_{\mu \nu} \rangle$. The relevant diagrams are depicted in
Figs. 2 and 3. As stated in Sec. I, our analysis goes from the
\lq\lq diagrammatic point of view'': We apply the HTL-resummation
scheme \cite{pis1,bra1,bra2,fre1} to each diagram in Figs. 2 and 3,
and extract the contributions at high-$T$, which are of the same
order as those from the lowest-order diagrams, Fig. 1.

In each diagram in Figs. 2 and 3, the part enclosed with the box is
the one-loop correction to the composite vertex, which we write
$\langle P | \Theta_{\mu \nu} | P \rangle$. We shall see that, for
hard $K \; (\sim T)$ and soft $P \; (\sim g T)$ [cf. Figs. 2 and 3],
some of these matrix elements are HTL's, i.e., are of the same order
as the lowest-order counterparts, and some of them lead to the
same-order contributions to the energy and pressure densities as
those from the lowest-order diagrams.

The HTL's summarize the leading thermal effect and are essentially
classical \cite{kel}. Then, in particular, they are free from
UV divergences, so that we do not need to take the
operator-mixing problem into consideration. In this relation, it is
worth recalling here that the $T$-dependent parts of the
above-mentioned $\langle P | \Theta_{\mu \nu} | P \rangle$ represent
\cite{bra1,fre1,eff-act2} essentially the matrix elements in tree
approximation in {\em vacuum theory}. In order to see this in the
present case, we take massless scalar $\phi^3$ theory, for
simplicity, and consider the composite operator $\phi^2 / 2$. We
compute one-loop correction to the composite vertex as depicted in
Fig. 4:
\begin{equation}
\langle p | \, \frac{1}{2} \, \phi^2 (0) \, | p \rangle \equiv -
\frac{g^2}{2} \;
\mbox{Tr}
\raisebox{-2.6mm}{\scriptsize{$\!\!\!\!\!\!\!$} K} \left\{ \Delta^+
(K) \right\}^2 \Delta^+ ( K + P ) \, .
\label{rei0}
\end{equation}
The manipulation of (\ref{rei0}) is carried out in Appendix B. After
continuing $p_0 \; (= 2 \pi i n T)$ to real energy $p_0 + i 0^+$, we
have (cf. (\ref{rei-3}))
\begin{eqnarray}
\langle p | \, \frac{1}{2} \, \phi^2 (0) \, | p \rangle
& = & \frac{1}{2} \, g^2 \int
\frac{d^{\, 4} K}{(2 \pi)^4} \sum_{i, \, j = 1}^2 (-)^{i + j}
D_{1 i} (K) D_{i j} (K) \nonumber \\
& & \mbox{\hspace*{18.1ex}} \times D_{j 1} (K + P) \, ,
\label{real-time}
\end{eqnarray}
the formula which is written in terms of the real-time formulation
of thermal field theory \cite{lan}, which is formulated on the time
path $- \infty$ $\to$ $+ \infty$ $\to$ $- \infty$ $\to$ $- \infty -
i / T$, in a complex time plane. In (\ref{real-time}), $P$ and $K$
are the four-vectors in the Minkowski space, \lq\lq 1'',
\lq\lq$i$'', and \lq\lq $j$'' are the thermal indices, and $D_{i j}$
$(i, j = 1, 2)$ is the $(i, j)$-component of the matrix propagator
of a massless scalar particle, whose explicit form is given, e.g.,
in \cite{lan,nie-sem}.

A thermal amplitude essentially represents the sum of the
corresponding matrix elements $\sum_{i, f} \langle i | S | f \rangle
\langle f | S^* | i \rangle$, where $S$ is the $S$-matrix in
{\em vacuum} theory, $S^*$ is the complex conjugate of the
$S$-matrix, and $| i \rangle$ [$| f \rangle$] is the initial [final]
state, which includes all the particles in the heat bath. The
details and the rules for the correspondences are presented in
\cite{nie}. We can apply the rules to each term in
(\ref{real-time}). As an example, we take (\ref{real-time}) with $i
= 1$ and $j = 2$ with $k_0 > 0$ and $k_0 + p_0 > 0$:
\begin{eqnarray}
& & \frac{g^2}{8 \pi^2} \int d^{\, 4} K \Bigg[ \left\{ \frac{1}{K^2
+ i 0^+} \, \delta (K^2) \otimes \delta ((K + P)^2) \right\} n_B (k)
\left( 1 + n_B (E) \right) \nonumber \\
& & \mbox{\hspace*{15ex}} - \left\{ 2 \pi i \, \delta^2 (K^2)
\otimes \delta ((K + P)^2) \right\} n_B^2 (k) (1 + n_B (E) ) \Bigg]
\, ,
\label{rei-4}
\end{eqnarray}

\noindent where $E = | {\bf k} + {\bf p} |$. It should be remembered
\cite{lan,nie-sem} that, although products of singular functions
appear in (\ref{rei-4}), they are cancelled in (\ref{real-time}),
leaving well-defined terms. The first and second terms in
(\ref{rei-4}) represent the products of {\em tree amplitudes}, as
depicted in Figs. 5(a) and (b), respectively. In Fig. 5, the
left-side part of the final-state cut line (dot-dashed line)
represents the amplitude or $S$-matrix element in {\em vacuum
theory}, while the right-side part represents the complex conjugate
of the amplitude. The state at the left end and the state at the
right end are the same and they represent the initial state. The
lines on top of Fig. 5 stand for the spectator particles, which are
the constituents of the heat bath. The left-side part of $\otimes$
in the first [second] curly brackets in (\ref{rei-4}) stands for the
amplitude $S$ in Fig. 5(a) [(b)] and the right-side part stands for
complex conjugate of the amplitude, $S^*$, in Fig. 5(a) [(b)]. Then,
each term in (\ref{rei-4}) is simply an integration over ${\bf k}$
of the product of the tree amplitude and the complex conjugate of
the tree amplitude weighted with statistical factors. Since the
latter factors dump at the UV region, no UV disaster arises. This
observation applies to all other portions of (\ref{real-time}),
provided that the $T = 0$ part has been subtracted.

The above argument applies to the $T$-dependent parts of all
$\langle P | \Theta_{\mu \nu} |P \rangle$'s in Figs. 2 and 3. Since
no UV divergences appear, $\langle P | \Theta_{0 \mu}|P \rangle$
[$\langle P | \Theta_{1 1}|P \rangle$] enjoys a meaning of
energy-momentum [pressure].

To summarize, as far as the $T$-dependent part of $\langle
\Theta_{\mu \nu} \rangle$, which yields leading-order contributions
at high-$T$, is concerned, we do not need to worry about the
operator-mixing problem, and we can use the operators
(\ref{theta-g}) and (\ref{theta-q}) for evaluating the
energy-momentum density.

Now we separately study the soft-gluon case, Fig. 2, and the
soft-quark case, Fig. 3.
\subsection{Soft-gluon sector}
In this subsection we extract from formally ${\cal O} (g^2)$
diagrams, Fig. 2, the contributions which are of the same order as
those from the tadpole diagrams Figs. 1(a) and (b), Eqs.
(\ref{g0}) - (\ref{g0-2}).

The contribution of Fig. 2(a) is written as
\begin{eqnarray}
\langle \Theta_{\mu \nu} \rangle
\rule[-3mm]{.14mm}{7mm} \raisebox{-2mm}{\scriptsize{$\;$Fig. 2(a)}}
& = & \frac{g^2}{2} N \left( N^2 - 1 \right) \; \mbox{Tr}
\raisebox{-2.6mm}{\scriptsize{$\!\!\!\!\!\!\!\!\!\!$soft}
P} \; \mbox{Tr} \raisebox{-2.6mm}{\scriptsize{$\!\!\!\!\!\!\!$} K}
\left\{ \Delta^+ (K) \right\}^2 \Delta^+
(K + P) \nonumber \\
& & \times
\displaystyle{\raisebox{0.6ex}{\scriptsize{*}}} \!
\Delta_{\gamma \delta} (P) \, \theta^{\alpha \beta}_{\mu \nu} (K)
{\cal V}_{\beta \gamma \epsilon} (K, P) \, \,
{\cal V}^\epsilon_{\,\; \delta \alpha} (K + P, - P) \, ,
\label{fig2a1}
\end{eqnarray}
\noindent where
\[
{\cal V}_{\beta \gamma \epsilon} (K, P) = g_{\beta \gamma}
(P - K)_\epsilon - g_{\gamma \epsilon} (K + 2 P)_\beta +
g_{\epsilon \beta} (2 K + P)_\gamma \, .
\]
In (\ref{fig2a1}), $\displaystyle{\raisebox{0.6ex}{\scriptsize{*}}}
\! \Delta_{\gamma \delta} (P)$ is as in (\ref{g-eff-pro}) and
$\theta^{\alpha \beta}_{\mu \nu} (K)$ in (\ref{2-pt-theta}).
Carrying out the contractions with respect to the repeated indices
and changing the integration variable $K$ to $ - K - P$ at several
places, (\ref{fig2a1}) is led to a sum of terms, which are of the
following generic forms,
\begin{eqnarray}
& & \mbox{Tr}
\raisebox{-2.6mm}{\scriptsize{$\!\!\!\!\!\!\!\!\!\!$soft} P} \;
\frac{1}{P^2 - \Pi_\ell (P)} \, F (P, K) \otimes I^{+ +}_{\, i \, j}
\, , \;\; (\ell = T, L) \, , \\
& & \mbox{Tr}
\raisebox{-2.6mm}{\scriptsize{$\!\!\!\!\!\!\!\!\!\!$soft} P} \;
\frac{1}{P^{2 n}} \, \, G (P, K) \otimes I^{+ +}_{\, i \, j} \, ,
\;\; (n = 1, 2) \, ,
\label{bonnrei22}
\end{eqnarray}
where
\begin{eqnarray}
F (P, K) \otimes I^{r r'}_{i \, j} & \equiv & \mbox{Tr}
\raisebox{-2.6mm}{\scriptsize{$\!\!\!\!\!\!\!$} K} \; F (P, K)
\left\{ \Delta^r ( K ) \right\}^i \nonumber \\
& & \mbox{\hspace*{3ex}} \times \left\{ \Delta^{r'} ( K + P )
\right\}^j \, , \;\;\;\; (r, \, r' = \pm) \, .
\label{hinagata}
\end{eqnarray}
Keeping the terms which lead to the contributions of the order
under consideration, we have
\begin{equation}
\langle \Theta_{\mu \nu}^{(g)} \rangle
\rule[-3mm]{.14mm}{7mm} \raisebox{-2mm}{\scriptsize{$\;$gluon}}
\simeq g^2 N (N^2 - 1) \; \mbox{Tr}
\raisebox{-2.6mm}{\scriptsize{$\!\!\!\!\!\!\!\!\!\!$soft} P}
\Bigg[ \sum_{\ell = T, \, L} \frac{1}{P^2 - \Pi_\ell (P)}
\, {\cal V}_{\mu \nu}^{(\ell)} + \frac{1}{P^4} \,
{\cal V}_{\mu \nu}^{(4)} + \frac{1}{P^2} \, {\cal V}_{\mu \nu}^{(2)}
\Bigg] \, ,
\label{i}
\end{equation}
where
\begin{eqnarray}
& & {\cal V}_{\mu \nu}^{(\ell)} =
\left( c_1^{(\ell)} g_{\mu \nu} + c_2^{(\ell)}
{\cal Q}_{\mu \nu}^{(\ell)} \right) I^+_{\, 1 \, 0} \nonumber \\
& & \mbox{\hspace*{6.8ex}}
+ \left( c_3^{(\ell)} g_{\mu \nu} \left(K
| {\cal Q}^{(\ell)} | K \right) + c_4^{(\ell)} \left( K_\mu
{\cal Q}^{(\ell)}_{\nu \rho} + K_\nu {\cal Q}^{(\ell)}_{\mu \rho}
\right) K^\rho + c_5^{(\ell)} K_\mu K_\nu \right) \otimes
I^+_{\, 2 \, 0} \nonumber \\
& & \mbox{\hspace*{6.8ex}}
+ \left( c_6^{(\ell)} g_{\mu \nu} \left( K |{\cal Q}^{(\ell)}| K
\right) + c_7^{(\ell)} \left( K_\mu {\cal Q}^{(\ell)}_{\nu \rho} +
K_\nu {\cal Q}^{(\ell)}_{\mu \rho} \right) K^\rho + c_8^{(\ell)}
K_\mu K_\nu \right) \otimes I^{+ +}_{\, 1 \, 1} \nonumber \\
& & \mbox{\hspace*{6.8ex}} + c_9^{(\ell)} K_\mu K_\nu \left( K
|{\cal Q}^{(\ell)}| K \right) \otimes I^{+ +}_{\, 2 \, 1} \, ,
\label{general1} \\
& & {\cal V}_{\mu \nu}^{(4)} = d_1 P_\mu P_\nu I^+_{\, 1 \, 0} +
P \cdot K \left\{ d_2 (K_\mu P_\nu + K_\nu P_\mu) + d_3
(P \cdot K) g_{\mu \nu} \right\} \otimes I^+_{\, 2 \, 0} \, , \\
& & {\cal V}_{\mu \nu}^{(2)} = d_4 \, g_{\mu \nu} I^+_{1 0} + d_5 \,
K_\mu K_\nu \otimes I^+_{\, 2 \, 0} + d_6 \, K_\mu K_\nu \otimes
I^{+ +}_{\, 1 \, 1} \, .
\label{22d}
\end{eqnarray}

\noindent In (\ref{i}) - (\ref{22d}), ${\cal Q}_{\mu \nu}^{(\ell)}$
is as in (\ref{lon-vac}) and
\begin{eqnarray}
& & \left(K | {\cal Q}^{(\ell)} | K \right) \equiv K^\rho \,
K^\sigma \, {\cal Q}_{\rho \sigma}^{(\ell)} \, ,
\label{q} \\
& & c_j^{(L)} = c_j^{(T)} \;\;\;\;\;\; \mbox{for} \;\; j = 2, 3, 4,
6, 7, 9 \, ,
\label{coeff} \\
& & c_j^{(L)} = \frac{1}{2} \, c_j^{(T)} \;\;\;\;\;\; \mbox{for}
\;\; j = 1, 5, 8 \, .
\label{cT-vs-cL}
\end{eqnarray}
The coefficients, $c$'s and $d$'s, in (\ref{i}) - (\ref{22d}) are
listed in the first row of Table I.

In a similar manner, we can show that the contributions of Figs.
2(b) - (f) may also be written in the form (\ref{i}) - (\ref{22d})
with (\ref{coeff}) and (\ref{cT-vs-cL}). [The dashed-lines in Figs.
2(e) and (f) stand for the hard FP-ghost propagators.] For each
contribution, the coefficients are tabulated in Table I.

Now we turn to analyze Figs. 2(g) - (i), where the dashed lines
stand for the FP-ghost propagators. One ghost-gluon vertex is
proportional to the soft ghost momentum. Then, this ghost-gluon
vertex brings in one power of a soft momentum, instead of a hard
loop momentum as in the case of Figs. 2(a) - (f). Then, the
contributions of Figs. 2(g) - (i) are negligible,
\begin{equation}
\langle \Theta_{\mu \nu}^{(\mbox{\scriptsize{ghost}})} \rangle
\simeq 0 \, .
\label{g-result1}
\end{equation}

Finally, we analyze Figs. 2(j) and (k). It can easily be shown
that, adding the contributions from Figs. 2(j) and (k), the
\lq\lq $g_{\mu \nu}$ term'' in (\ref{theta-q}) --- the term that
is proportional to $g_{\mu \nu}$ --- leads to vanishing
contribution. It is worth pointing out in passing that the $g_{\mu
\nu}$ term vanishes provided the equation of motion is used. Then,
what we should analyze is Figs. 2(j) and (k) with the first term on
the r.h.s. of (\ref{theta-q}). It is an easy task to see that the
term $ P_\alpha P_\beta / P^4$ of the effective soft-gluon
propagator $\displaystyle{\raisebox{0.6ex}{\scriptsize{*}}} \!
\Delta_{\alpha \beta} (P)$ (cf. (\ref{g-eff-pro})) does not yield
leading contribution. Ignoring all other terms that lead to
nonleading contributions, we obtain for the contribution to $\langle
\Theta_{\mu \nu}^{(g)} \rangle$,
\begin{eqnarray}
\langle \Theta_{\mu \nu}^{(g)} \rangle
\rule[-3mm]{.14mm}{7mm} \raisebox{-2mm}{\scriptsize{$\;$quark}}
& = & 2 g^2 N_f \left( N ^2 - 1 \right) \;\,
\mbox{Tr} \raisebox{-2.6mm}{\scriptsize{$\!\!\!\!\!\!\!\!\!\!$soft}
P} \Bigg[ \sum_{\ell = T, \, L}
\frac{1}{ P^2 - \Pi_\ell (P)} \left\{ {\cal Q}^{(\ell)}_{\mu \nu}
\otimes I^-_{1 0} \right. \nonumber \\
& & \left. + \hat{c}^{(\ell)} K_\mu K_\nu \otimes I^-_{\, 2 \, 0}
- 2 \left( K_\nu
{\cal Q}_{\mu \sigma}^{(\ell)} + K_\mu
{\cal Q}_{\nu \sigma}^{(\ell)} \right) K^\sigma \otimes
I^{- -}_{\, 1 \, 1} \right. \nonumber \\
& & \left. + 4 K_\mu K_\nu \left( K |
{\cal Q}^{(\ell)} | K \right) \otimes I^{- -}_{\, 2 \, 1} \right\}
\Bigg] \, ,
\label{glu-quar}
\\
& & \hat{c}^{(T)} = 2 \hat{c}^{(L)} = - 2
\nonumber \, .
\end{eqnarray}

\noindent In (\ref{glu-quar}), $I^{- -}_{\, i \, j}$ etc. are
defined as in (\ref{hinagata}).
\subsection{Soft-quark sector}
In this subsection we extract from formally ${\cal O} (g^2)$
diagrams the contributions which are of the same order as that from
the tadpole diagram Fig. 1 (c), Eq. (\ref{q01}).

The diagrams to be analyzes are depicted in Figs. 3(a) - (c). The
leading contribution from Figs. 3(a) and (b) is obtained as
\begin{eqnarray}
\langle \Theta_{\mu \nu}^{(q)} \rangle
\rule[-3mm]{.14mm}{7mm} \raisebox{-2mm}{\scriptsize{$\;$quark}}
& \simeq & 2 g^2 \, N_f
\left( N^2 - 1 \right) \nonumber \\
& & \times \; \mbox{Tr}
\raisebox{-2.6mm}{\scriptsize{$\!\!\!\!\!\!\!\!\!\!$soft} P}
\sum_{\sigma = \pm} \displaystyle{\raisebox{0.6ex}{\scriptsize{*}}}
\! \tilde{S}^{(\sigma)} (P) \left[ 4 K_\mu K_\nu \left( K \cdot
\hat{P}_\sigma \right) \otimes I^{- +}_{\, 2 \, 1}
\right. \nonumber \\
& & \left. + \left(
\hat{P}_{\sigma \mu} K_\nu + \hat{P}_{\sigma \nu} K_\mu \right)
\otimes I^{- +}_{\, 1 \, 1} \right] \, ,
\label{3ab}
\end{eqnarray}
while Fig. 3(c) yields
\begin{eqnarray}
\langle \Theta_{\mu \nu}^{(q)} \rangle
\rule[-3mm]{.14mm}{7mm} \raisebox{-2mm}{\scriptsize{$\;$gluon}}
& \simeq & - 4 g^2 N_f (N^2 - 1) \; \mbox{Tr}
\raisebox{-2.6mm}{\scriptsize{$\!\!\!\!\!\!\!\!\!\!$soft} P}
\sum_{\sigma = \pm} \displaystyle{\raisebox{0.6ex}{\scriptsize{*}}}
\! \tilde{S}^{(\sigma)} (P) \left[ 2 K_\mu K_\nu \left( K \cdot
\hat{P}_\sigma \right) \otimes I^{+ -}_{\, 2 \, 1} \right.
\nonumber \\
& & \mbox{\hspace*{35.9ex}} \left. - \left( K_\mu
\hat{P}_{\sigma \nu} + K_\nu \hat{P}_{\sigma \mu} \right) \otimes
I^{+ -}_{\, 1 \, 1} \right]
\, . \nonumber \\
\label{3c}
\end{eqnarray}
\section{Energy density of soft modes}
In this section we compute the energy ${\bf P}_0$, Eq.
(\ref{P-mu}), of soft modes.

\subsection{Gluon sector}

\noindent {\em Tadpole diagrams, Figs. 1(a) and (b)}.

 From (\ref{g0}) -  (\ref{lon-vac}), it is straightforward to
evaluate the energy of soft modes, ${\bf P}_0^{(g 0)}$ (cf.
(\ref{P-mu})). Employing the spectral representation
(\ref{spect1}) in Appendix A for $\left( P^2 - \Pi_\ell (P)
\right)^{- 1}$ in (\ref{g-eff-pro}) and using (\ref{tau}) in
Appendix C for $\Delta^+ (P)$ in (\ref{g0-2}), we obtain
\begin{eqnarray*}
\langle \Theta_{0 0}^{(g 0)} \rangle & = & \frac{1}{2} \left( N^2
- 1 \right) \; \mbox{Tr}
\raisebox{-2.6mm}{\scriptsize{$\!\!\!\!\!\!\!\!\!\!$soft} P}
\int_0^{1 / T} d \tau \, e^{p_0 \tau}
\int_{- \infty}^{+ \infty} d \xi \left( 1 + n_B (\xi) \right)
e^{- \xi \tau}
\\
& & \times \left[ 2 \, \rho_T (\xi) \left( p_0^2 + p^2 \right) +
\rho_L (\xi) \left( p_0^2 - p^2 \right) \right]
\\
& & - \frac{1}{2} \left( N^2 - 1 \right) \; \mbox{Tr}
\raisebox{-2.6mm}{\scriptsize{$\!\!\!\!\!\!\!\!\!\!$soft} P} \, 1
\, .
\end{eqnarray*}

\noindent where
\begin{equation}
n_B (\xi) = \frac{1}{e^{\xi / T} - 1} \, ,
\label{bose}
\end{equation}
and the spectral functions, $\rho$'s, are defined as in
(\ref{spect3}). Summation over $n$ (cf. (\ref{symbol})) and
integration over $\tau$ lead to
\begin{eqnarray}
\langle \Theta_{0 0}^{(g 0)} \rangle & = & \frac{1}{2} \left( N^2
- 1 \right) \int \frac{d^{\, 3} p}{(2 \pi)^3} \int_{- \infty}^{+
\infty} d \xi \, n_B (\xi) \left\{
2 (\xi^2 + p^2) \, \rho_T (\xi) + (\xi^2 - p^2 ) \, \rho_L (\xi)
\right\} \nonumber \\
& & - 2 \left( N^2 - 1 \right) \; \mbox{Tr}
\raisebox{-2.6mm}{\scriptsize{$\!\!\!\!\!\!\!\!\!\!$soft} P} \, 1
\, .
\label{g02}
\end{eqnarray}

\noindent Here and in the following, unless otherwise indicated,
integration over ${\bf p}$ is carried out over the region
({\ref{soft-region}}). In deriving (\ref{g02}), use has been made
of the spectral sum rule (\ref{sum-rule1}) in Appendix A. It is to
be noted that, in ${\bf P}^{(g 0)}_0$ (cf. (\ref{P-mu})), the last
term in (\ref{g02}) is cancelled by the term coming from
$\langle 0 | \Theta_{0 0}^{(g 0)}| 0 \rangle$. We recall that we are
interested in the region $p \leq {\cal O} (g T)$ and note that
$\rho_{T / L} (\xi)$ vanishes \cite{gluon} in the region $| \xi | >
p + {\cal O} (g T)$, where the term of ${\cal O} (g T)$ comes from
the pole of $\left( P^2 - \Pi_\ell (P) \right)^{- 1}$. Then, in the
region of our interest, $\xi / T << 1$, and we may approximate
$n_B (\xi)$ as
\begin{equation}
n_B (\xi) \simeq \frac{T}{\xi} \, .
\label{bose-app}
\end{equation}
Using the sum rules, (\ref{sum-rule1}) and (\ref{sum-rule2}) in
Appendix A, we obtain
\begin{eqnarray}
{\bf P}_0^{(g 0)} & = & \langle \Theta_{0 0}^{(g 0)} \rangle -
\langle 0 | \Theta_{0 0}^{(g 0)} | 0 \rangle
\label{g022} \\
& \simeq & (N^2 - 1) \int \frac{d^{\, 3} p}{(2 \pi)^3} \left[
\left\{ 2 T + \frac{3}{2} \, \frac{m_g^2 \, T}{p^2 + 3 \, m_g^2}
\right. \right. \nonumber \\
& & \left. \left. \mbox{\hspace*{20.5ex}} +
{\cal O} \left( \frac{p^2}{T} \right) \right\} - p \, \right] \, ,
\label{g03}
\end{eqnarray}
where $m_g$ is the effective gluon mass induced by the quark-gluon
plasma,
\begin{equation}
m_g^2 = \left( N + \frac{N_f}{2} \right) \frac{g^2 T^2}{9} \, .
\label{gluon-mass}
\end{equation}
In (\ref{g03}), $- p$ in the square brackets has come from $-
\langle 0 | \Theta_{0 0}^{(g 0)} | 0 \rangle$ in (\ref{g022}). If we
were to use the bare thermal gluon propagator $g_{\alpha \beta} /
P^2$ instead of $\displaystyle{\raisebox{0.6ex}{\scriptsize{*}}} \!
\Delta_{\alpha \beta} (P)$ in (\ref{g0-1}), we would obtain
(\ref{g03}) with the second term in the square brackets being
absent. It is to be noted in passing that (\ref{g03}) is valid for
$T >> p$. Then, it is of no surprise that (\ref{g03}) does not
vanish in the naive limit $T \to 0$.

\noindent {\em Figure 2.}

Figs. 2(a) - (f): The starting formulae are (\ref{i}) -
(\ref{cT-vs-cL}) with Table I. All the necessary formulae for
evaluating $\langle \Theta_{00} \rangle$ are displayed in Appendix
C. The result may be written in the form,
\begin{eqnarray}
& & \langle \Theta_{0 0}^{(g)} \rangle \rule[-3mm]{.14mm}{7mm}
\raisebox{-2mm}{\scriptsize{$\;$gluon}} \nonumber \\
& & \mbox{\hspace*{3ex}} \simeq g^2 N (N^2 - 1) \, T \int
\frac{d^{\, 3} p}{(2 \pi)^3} \Bigg[ \frac{7}{16} \, \frac{T^2}{p^2}
- \frac{7}{16} \, \frac{T^2}{p^2 + 3 m_g^2} \Bigg] \, , \nonumber \\
& & \mbox{\hspace*{3ex}} = \frac{21}{16} \, g^2 \, N (N^2 - 1) \, T
\int \frac{d^{\, 3} p}{(2 \pi)^3} \, \frac{m_g^2 \, T^2}{p^2 (p^2 +
3 m_g^2)} \, ,
\label{g-result}
\end{eqnarray}
which comes from small parts of the integration region in
(\ref{fig2a1}), where $K$ is hard, $k = {\cal O} (T)$ [HTL]. In
(\ref{g-result}), the contribution to the coefficients $7/16$
$(- 7/16)$ from Figs. 2(a) - (f) are, in respective order, as
follows,
\begin{eqnarray*}
& & \frac{7}{16} = \frac{49}{48} - \frac{11}{24} + \frac{1}{8} -
\frac{3}{16} - \frac{3}{16} + \frac{1}{8} \, , \\
& & - \frac{7}{16} = - \frac{1}{16} - \frac{1}{8} - \frac{1}{8} -
\frac{3}{16} + \frac{1}{48} + \frac{1}{24} \, .
\end{eqnarray*}

Figs. 2(g) - (i): As discussed in Sec. IV, in conjunction with
(\ref{g-result1}), the contributions are nonleading as compared to
(\ref{g03}), in which the first two terms in the curly brackets are
kept.

Figs. 2(j) - (k): The relevant formula is (\ref{glu-quar}). Using
(\ref{fermion}) in Appendix C together with (\ref{koushiki1}) -
(\ref{koushiki2}), we obtain
\begin{eqnarray}
& & \langle \Theta_{0 0}^{(g)} \rangle
\rule[-3mm]{.14mm}{7mm} \raisebox{-2mm}{\scriptsize{$\;$quark}}
\nonumber \\
& & \mbox{\hspace*{4ex}} \simeq - \frac{1}{4} g^2 \, N_f \left(
N^2 - 1 \right) T \int \frac{d^{\, 3} p}{(2 \pi)^3} \,
\frac{T^2}{p^2 + 3 \, m_g^2} \, .
\label{g-result2}
\end{eqnarray}

Adding all the contributions (\ref{g03}), (\ref{g-result}),
(\ref{g-result1}), and (\ref{g-result2}), we obtain for the leading
contribution to ${\bf P}_0^{(g)}$,
\begin{eqnarray}
{\bf P}_0^{(g)} & = & \langle \Theta_{0 0}^{(g)} \rangle -
\langle 0 | \Theta_{0 0}^{(g)} | 0 \rangle \nonumber \\
& \simeq & \left( N^2 - 1 \right) T \int
\frac{d^{\, 3} p}{(2 \pi)^3} \left[ 2 + \frac{7}{16} N \left(
\frac{g T}{p} \right)^2 \right. \nonumber \\
& & \left. + \frac{1}{2 \left( p^2 + 3 \, m_g^2
\right)} \left\{ 3 \, m_g^2 - \frac{1}{2} \, \left( \frac{7}{4} N +
N_f \right) (g T)^2 \right\} \right] \nonumber \\
& = & \int_{\mbox{\scriptsize{soft}} \; p}
\frac{d^{\, 3} p}{(2 \pi)^3}
\, {\cal E}_g (p) \, ,
\label{g-final} \\
{\cal E}_g (p) & = & \left( N^2 - 1 \right) T \left[ 2 +
\frac{63}{8} \, \frac{N}{2 N + N_f} \left( \frac{m_g}{p} \right)^2
- \, \frac{3}{8} \, \frac{13 N + 8 N_f}{2 N + N_f} \,
\frac{m_g^2}{p^2 + 3 \, m_g^2} \right] \, ,
\label{g-final1}
\end{eqnarray}

\noindent where we have used (\ref{gluon-mass}). It is an elementary
task to show that ${\cal E}_g (p)$ is positive definite for any
values of $N$, $N_f$, $T$, and $p$.
\subsection{Quark sector}

\noindent {\em Tadpole diagram, Fig. 1(c). }

The starting formulae are (\ref{q01}) with $\mu = \nu = 0$ and
(\ref{S-hat}). Using the spectral representation (\ref{spect2}) in
Appendix A, summing over $n$ (cf. (\ref{symbol})), and carrying out
the integration over $\tau$, we have
\begin{eqnarray}
\langle \Theta_{0 0}^{(q 0)} \rangle & = & - 4 N_f \int
\frac{d^{\, 3} p}{(2 \pi)^3} \, p \int d \xi
\{ 1 - n_F (\xi) \} \nonumber \\
& & \mbox{\hspace*{17.6ex}} \times \left[ \sum_{\sigma = \pm} \sigma
\rho_\sigma (\xi) \right] \, ,
\label{q-energy}
\end{eqnarray}
where
\begin{equation}
n_F ( \xi ) = \frac{1}{e^{\xi / T} + 1} \, ,
\label{fermi}
\end{equation}
and the spectral function $\rho_\sigma$ is defined as in
(\ref{spect4}). Since $\rho_\sigma (\xi)$ is nonzero only in the
region $| \xi | \leq p + {\cal O} (g T)$, we may approximate
$n_F (\xi)$ as
\begin{equation}
n_F (\xi) \simeq \frac{1}{2} - \frac{1}{4} \, \frac{\xi}{T} + ...
\, .
\label{fermi-app}
\end{equation}
Then, using the sum rules (\ref{sum-rule3}) - (\ref{sum-rule6}) in
Appendix A, we obtain
\begin{equation}
\langle \Theta_{0 0}^{(q 0)} \rangle = N_f
\int_{\mbox{\scriptsize{soft}} \; p}
\frac{d^{\, 3} p}{(2 \pi)^3} \, \frac{p^2}{T} \, \left[ 1 + {\cal O}
\left( \frac{p^2 + 5 m_f^2 / 3}{T^2} \right) \right] \, ,
\label{kin1}
\end{equation}
where
\begin{equation}
\mbox{\hspace*{-21ex}} m_f^2 = \frac{1}{16} \, \frac{N^2 - 1}{N} \,
(g T)^2 \, .
\label{quark-mass}
\end{equation}
Note that the effect of the HTL resummation for the quark propagator
appears at the term of ${\cal O} (m_f^2 / T^2) = {\cal O} (g^2)$
relative to 1. In fact, (\ref{q01}) with $S (P)$ in place of
$\displaystyle{\raisebox{0.6ex}{\scriptsize{*}}} \! S (P)$ leads to
(\ref{q-energy}) in which $\rho_\sigma (\xi) = \frac{1}{2} \delta
(\xi - \sigma p)$:
\begin{eqnarray}
\langle \Theta_{0 0}^{(q 0)} \rangle \rule[-3.3mm]{.14mm}{8mm}
\raisebox{-3mm}{\scriptsize{$
\displaystyle{\raisebox{0.6ex}{\scriptsize{*}}} \! S
\to S$}} & = & - 2 N_f \int \frac{d^{\, 3} p}{(2 \pi)^3} \, p
\sum_{\sigma = \pm} \sigma \, \{ 1 - n_F (\sigma p) \} \nonumber \\
\label{kin2} \\
& \simeq & N_f
\int \frac{d^{\, 3} p}{(2 \pi)^3} \, \frac{p^2}{T} \left[ 1 +
{\cal O} \left( \frac{p^2}{T^2} \right) \right] \, .
\label{kin3}
\end{eqnarray}

Now we rewrite (\ref{kin2}) as
\begin{eqnarray}
\mbox{(\ref{kin2})} & = & 4 N_f \int \frac{d^{\, 3} p}{(2 \pi)^3} \,
\left\{ p \, n_F (p) - \frac{p}{2} \right\}
\label{kin5} \\
& = & 4 N_f \int \frac{d^{\, 3} p}{(2 \pi)^3} \, p \, n_F (p) +
\langle 0 | \Theta_{0 0 }^{(q 0)} | 0 \rangle \, .
\label{kin6}
\end{eqnarray}
Using (\ref{kin1}), (\ref{kin3}), and (\ref{kin6}), we obtain
\begin{eqnarray}
{\bf P}_0^{(q 0)} & = & \langle \Theta_{0 0}^{(q 0)} \rangle -
\langle 0 | \Theta_{0 0}^{(q 0)} | 0 \rangle
\label{q033} \\
& \simeq & \int_{\mbox{\scriptsize{soft}} \; p}
\frac{d^{\, 3} p}{(2 \pi)^3} \, {\cal E}_q (p) \, ,
\label{q04} \\
{\cal E}_q (p) & = & N_f \left[
2 p + {\cal O} \left( \frac{p^2 + 5 m_f^2 / 3}{T} \right) \right]
\, .
\label{q044}
\end{eqnarray}
In contrast to the case of gluon sector, the $T$-dependent part of
${\cal E}_q (p)$, Eq. (\ref{q044}), is nonleading. The origin of
this difference may be traced back to the different statistics,
(\ref{bose-app}) and (\ref{fermi-app}).

Here some comments are in order. The factor \lq\lq 4'' in
(\ref{kin5}) - (\ref{kin6}) represents four degrees of freedom,
$4 = 2 \times 2$,
where the first \lq\lq 2'' comes from a quark and an antiquark and
the second \lq\lq 2'' from spin degrees of freedom. $p n_F (p)$ in
(\ref{kin5}) represents the thermal energy per one mode of momentum
${\bf p}$ and $ - p / 2$ is the zero-point energy. In fact the
zero-point energy per one fermionic mode of momentum ${\bf p}$ is
obtained as $\langle 0 | \frac{p}{2} \, (a_{\bf p}^\dagger a_{\bf p}
- a_{\bf p} a_{\bf p}^\dagger) | 0 \rangle = - \frac{p}{2}$, where
$a_{\bf p}$ [$a_{\bf p}^\dagger$] is the annihilation [creation]
operator of an (anti)quark with ${\bf p}$ and with definite spin. It
is interesting to note that, for soft $p$, $n_F (p) \simeq 1 / 2$
(cf. (\ref{fermi-app})) and then, in (\ref{kin5}), (the leading part
of) the thermal energy and the zero-point energy cancel out to yield
(\ref{kin3}). ${\cal E}_q (p)$ in (\ref{q044}) is the thermal energy
of (anti)quarks with momentum ${\bf p}$, $N_f \times 4 \times p
\times n_F (p) \simeq 2 N_f \times p$.

\noindent {\em Figure 3.}

Figs. 3(a) and (b): The contribution to the energy density is
obtained from (\ref{3ab}) as

\begin{eqnarray}
\langle \Theta_{0 0}^{(q)} \rangle
\rule[-3mm]{.14mm}{7mm} \raisebox{-2mm}{\scriptsize{$\;$quark}}
& \simeq & 4 g^2 N_f \left( N^2 - 1 \right) \nonumber \\
& & \times \; \mbox{Tr}
\raisebox{-2.6mm}{\scriptsize{$\!\!\!\!\!\!\!\!\!\!$soft} P}
\sum_{\sigma = \pm} \displaystyle{\raisebox{0.6ex}{\scriptsize{*}}}
\! \tilde{S}^{(\sigma)} (P) \left\{ 2 k^2 \left( K \cdot
\hat{P}_\sigma \right) \otimes I^{- +}_{\, 2 \, 1} + \left( 3 k_0 -
2 \sigma {\bf k} \cdot \hat{{\bf p}} \right) \otimes
I^{- +}_{\, 1 \, 1} \right\} \, . \nonumber \\
\label{q-1}
\end{eqnarray}

\noindent With the help of (\ref{spect2}) in Appendix A and
(\ref{tau}) in Appendix C, it is straightforward to show that the
$I^{- +}_{\, 1 \, 1}$ part in (\ref{q-1}) leads to a nonleading
contribution. As to the contribution with $I^{- +}_{\, 2 \, 1}$ part
in (\ref{q-1}), using (\ref{tau}) and (\ref{tau1}), the relevant HTL
parts are easily evaluated:
\begin{eqnarray}
& & T \sum_{n = - \infty}^{+ \infty}
\displaystyle{\raisebox{0.6ex}{\scriptsize{*}}} \!
\tilde{S}^{(\sigma)} (P) \, k^3 F (\hat{{\bf k}} \cdot
\hat{{\bf p}}) \otimes I^{- +}_{\, 2 \, 1} \nonumber \\
& &
\mbox{\hspace*{6ex}}
\simeq - \frac{1}{8 T}
\int_{- \infty}^{+ \infty} d \xi \, \rho_\sigma (\xi)
\int \frac{d^{\, 3} k}{(2 \pi)^3} \, F (\hat{\bf k} \cdot
\hat{{\bf p}}) \nonumber \\
& &
\mbox{\hspace*{9ex}}
\times \sum_{s = \pm} \left[ n_F (1 - n_F) (1 + 2 n_B) \,
\frac{s}{\hat{{\bf k}} \cdot {\bf p} - s \xi} \right.
\nonumber \\
& & \left.
\mbox{\hspace*{9ex}}
+ n_B (1 + n_B)
(1 - 2 n_F) \,
\frac{s \hat{{\bf k}} \cdot {\bf p}}{(\hat{{\bf k}} \cdot {\bf p} -
s \xi)^2} \right] \nonumber \\
& & \mbox{\hspace*{6ex}} = - \frac{7}{128 \pi^2} \, \zeta (3) \, T^2
\int d (\hat{{\bf k}} \cdot \hat{{\bf p}}) \, F (\hat{\bf k} \cdot
\hat{{\bf p}}) \nonumber \\
& & \mbox{\hspace*{9ex}} \times
\int_{- \infty}^{+ \infty} d \xi \, \rho_\sigma (\xi)
\sum_{s = \pm} \left[ \frac{s}{\hat{{\bf k}} \cdot {\bf p} - s \xi}
+ \frac{s \hat{{\bf k}} \cdot {\bf p}}{(\hat{{\bf k}} \cdot {\bf p}
- s \xi)^2} \right] \, ,
\label{fer-bos-1} \\
& & T \sum_{n = - \infty}^{+ \infty}
\displaystyle{\raisebox{0.6ex}{\scriptsize{*}}} \!
\tilde{S}^{(\sigma)} (P) \, k_0 \, k^2 \, G (\hat{{\bf k}} \cdot
\hat{{\bf p}}) \otimes I^{- +}_{\, 2 \, 1} \nonumber \\
& & \mbox{\hspace*{6ex}} \simeq - \frac{7}{128 \pi^2} \, \zeta (3)
\, T^2 \int d (\hat{{\bf k}} \cdot \hat{{\bf p}}) \, G (\hat{\bf k}
\cdot \hat{{\bf p}}) \nonumber \\
& & \mbox{\hspace*{9ex}} \times
\int_{- \infty}^{+ \infty} d \xi \, \rho_\sigma
(\xi) \sum_{s = \pm} \left[ \frac{1}{\hat{{\bf k}} \cdot {\bf p} - s
\xi} + \frac{\hat{{\bf k}} \cdot {\bf p}}{(\hat{{\bf k}} \cdot
{\bf p} - s \xi)^2} \right] \, .
\label{fer-bos-2}
\end{eqnarray}

\noindent We note that, in (\ref{q-1}),
\begin{equation}
K \cdot \hat{P}_\sigma = k_0 - \sigma {\bf k} \cdot \hat{{\bf p}}
\, .
\label{q-2}
\end{equation}
Eq. (\ref{q-1}) with the $k_0$ part in (\ref{q-2}) leads to
(\ref{fer-bos-2}) with $G (\hat{{\bf k}} \cdot \hat{{\bf p}}) = 1$,
which vanishes because the integrand is odd with respect to
$\hat{{\bf k}} \cdot \hat{{\bf p}}$. The other part in (\ref{q-2}),
$- \sigma {\bf k} \cdot \hat{{\bf p}}$, leads to (\ref{fer-bos-1})
with $F (\hat{{\bf k}} \cdot \hat{{\bf p}}) = - \sigma \hat{{\bf k}}
\cdot \hat{{\bf p}}$, which also vanishes due to the same reason as
above. Hence the contribution (\ref{q-1}) is of nonleading,
\[
\langle \Theta_{0 0}^{(q)} \rangle
\rule[-3mm]{.14mm}{7mm} \raisebox{-2mm}{\scriptsize{$\;$quark}}
\simeq 0 \, .
\]

Fig. 3(c): The starting formula is (\ref{3c}). The same analysis as
above shows that
\[
\langle \Theta_{0 0}^{(q)} \rangle
\rule[-3mm]{.14mm}{7mm} \raisebox{-2mm}{\scriptsize{$\;$gluon}}
\simeq 0 \, .
\]

In conclusion, to leading order at high-$T$, ${\bf P}^{(q 0)}_0$,
Eq. (\ref{q04}), does not receive additional contributions from
Figs. 3(a) - (c).
\section{Hydrostatic pressure}
In this section, we compute the hydrostatic or kinetic pressure
density of soft modes. The $T$-dependent part of the hydrostatic
pressure ${\bf P}$ is defined as in (\ref{Pressure}):
\begin{eqnarray}
{\bf P} & \equiv & \langle \Theta_{1 1} \rangle - \langle 0 |
\Theta_{1 1} | 0 \rangle \nonumber \\
& = & - \frac{1}{3} \left[ \left\{ \langle \Theta_\mu^{\;\, \mu}
\rangle - \langle 0 | \Theta_\mu^{\;\; \mu} | 0 \rangle \right\} -
\left\{ \langle \Theta_{0 0} \rangle - \langle 0 | \Theta_{0 0} | 0
\rangle \right\} \right] \, . \nonumber \\
\label{Pressure1}
\end{eqnarray}
We are interested in the leading contribution at high $T$ to
(\ref{Pressure1}) for soft modes. Then, as in Sec. V, no diverging
integral appears.

We shall show that, to leading order,
\begin{equation}
\langle \Theta_\mu^{\;\; \mu} \rangle - \langle 0 |
\Theta_\mu^{\;\; \mu} | 0 \rangle \simeq 0 \, .
\label{Pressure2}
\end{equation}
Using this in (\ref{Pressure1}), we obtain
\[
{\bf P} \simeq \frac{1}{3} \left\{ \langle \Theta_{0 0} \rangle -
\langle 0 | \Theta_{0 0} | 0 \rangle \right\} = \frac{1}{3} \,
{\bf P}_0 \, ,
\]
where ${\bf P}_0$ is as in (\ref{P-mu}) with $\mu = 0$. Then,
defining the \lq\lq pressure density'' ${\cal P} (p)$ as
\[
{\bf P} \equiv \int_{\mbox{\scriptsize{soft}} \; p}
\frac{d^{\, 3} p}{(2 \pi)^3} \, {\cal P} (p) \, ,
\]
we have, for the gluon [quark] sector, ${\cal P}_g (p) \simeq
{\cal E}_g (p) / 3$ [${\cal P}_q (p) \simeq {\cal E}_q (p) / 3$ ],
where ${\cal E}_g$ [${\cal E}_q$], Eq. (\ref{g-final1}) [Eq.
(\ref{q044})], is the energy of soft gluon [quark] with momentum
$p$.

We are now in a position to confirm (\ref{Pressure2}). From
(\ref{theta-g}) with $\lambda = 1$ and (\ref{theta-q}), we obtain
\begin{eqnarray}
& & \Theta_{\,\; \mu}^{(g) \mu} = 2 \left[
\left( \partial_\mu \, \bar{\omega}^a \right) \left( D^\mu \, \omega
\right)^a - \partial^\mu \left( A_\mu^a \, \partial \cdot A^a
\right) \right] \, ,
\label{g-trace} \\
& & \Theta_{\;\, \mu}^{(q) \mu} = - 3 i \, \bar{\psi}
\stackrel{\leftrightarrow}
{{D\kern-0.14em\raise0.17ex\llap{/}\kern0.15em\relax}} \, \psi \, .
\label{q-trace}
\end{eqnarray}
We note that a term with a total derivative does not contribute to
(\ref{Pressure2}). Then, the second term in the square brackets in
(\ref{g-trace}) may be ignored, and we may replace (\ref{g-trace})
by
\begin{equation}
\Theta_{\;\, \mu}^{(g) \mu} = - 2 \, \bar{\omega}^a \, \partial_\mu
\left( D^\mu \omega \right)^a \, .
\label{glu-tr}
\end{equation}

It is to be noted in passing that (\ref{q-trace}) and (\ref{glu-tr})
vanish when the equations of motion are imposed.

The diagrams to be analyzed are Figs. 1(b), Figs. 2(e), (f), (g),
and (h), with $\Theta_{\, \; \mu}^{(g) \mu}$ insertion, and Fig.
1(c), Figs. 2(j), (k), Figs. 3(a), and (b), with
$\Theta_{\,\; \mu}^{(q) \mu}$ insertion.

Fig. 1(b): We obtain for the contribution to $\langle
\Theta_{\,\; \mu}^{(q) \mu} \rangle$,
\begin{eqnarray}
\langle \Theta_{\,\; \mu}^{(g) \mu} \rangle
\rule[-3mm]{.14mm}{7mm} \raisebox{-2mm}{\scriptsize{$\;$Fig. 1(b)}}
& = & - (N^2 - 1) \,
\mbox{Tr} \raisebox{-2.6mm}{\scriptsize{$\!\!\!\!\!\!\!$} P} \;
\frac{1}{P^2} \cdot P^2 \nonumber \\
& = & - (N^2 - 1) \,
\mbox{Tr} \raisebox{-2.6mm}{\scriptsize{$\!\!\!\!\!\!\!$} P} \; \;
1 \, .
\label{1b}
\end{eqnarray}
As in (\ref{g02}) above, (\ref{1b}) is cancelled by $\langle 0 |
\Theta_{\,\; \mu}^{(g) \mu} | 0 \rangle$ in (\ref{Pressure2}):
\[
\left[ \langle \Theta_{\,\; \mu}^{(g) \mu} \rangle -
\langle 0 | \Theta_{\,\; \mu}^{(g) \mu} | 0 \rangle
\right]_{\mbox{\scriptsize{Fig. 1(b)}}} = 0 \, .
\]

Fig. 1(c): The contribution to $\langle \Theta_{\,\; \mu}^{(g) \mu}
\rangle$ is
\begin{equation}
\langle \Theta_{\,\; \mu}^{(q) \mu} \rangle
\rule[-3mm]{.14mm}{7mm} \raisebox{-2mm}{\scriptsize{$\;$Fig. 1(c)}}
= - 3 N_f \;\; \mbox{Tr}
\raisebox{-2.6mm}{\scriptsize{$\!\!\!\!\!\!\!\!\!\!$soft}
P} tr \left[ {P\kern-0.07em\raise0.17ex\llap{/}\kern0.15em\relax} \,
\displaystyle{\raisebox{0.6ex}{\scriptsize{*}}} \!
{S\kern-0.07em\raise0.17ex\llap{/}\kern0.15em\relax} (P) \right]
\, .
\label{1cyo}
\end{equation}
Computation of (\ref{1cyo}) goes through the similar procedure as in
Sec. VB. In place of (\ref{q-energy}), we have
\begin{eqnarray*}
\langle \Theta_{\,\; \mu}^{(q) \mu} \rangle & = & 12 N_f \int d \xi
\left\{ 1 - n_F (\xi) \right\} \nonumber \\
& & \mbox{\hspace*{7.8ex}} \times \sum_{\sigma = \pm} \int
\frac{d^{\, 3} p}{(2 \pi)^3} (\xi - \sigma p) \, \rho_\sigma (\xi)
\, .
\end{eqnarray*}
Then, using (\ref{fermi-app}) and (\ref{sum-rule3}) -
(\ref{sum-rule6}) in Appendix A, we obtain
\begin{eqnarray}
& & \left[ \langle \Theta_{\,\; \mu}^{(q) \mu} \rangle - \langle 0 |
\Theta_{\,\; \mu}^{(q) \mu} | 0 \rangle
\right]_{\mbox{\scriptsize{Fig. 1(k)}}} = 3
\int_{\mbox{\scriptsize{soft}} \; p} \frac{d^{\, 3} p}{(2 \pi)^3}
\, \theta_{\,\; \mu}^{(q) \mu}
(p) \, , \nonumber \\
\label{ro} \\
& & \theta_{\,\; \mu}^{(q) \mu} (p) \simeq N_f \, \frac{m_f^2}{T}
\, ,
\label{ha}
\end{eqnarray}

\noindent where $m_f$ is as in (\ref{quark-mass}).

\noindent In the region of our interest, Eq. (\ref{soft-region}),
(\ref{ro}) - (\ref{ha}) are subleading as compared to (\ref{q044}).

Figs. 2(e) and (f): It is straightforward to show that the
contribution of Fig. 2(e) cancels the contribution of Fig. 2(f).

Figs. 2(g) and (h): The contribution is
\begin{equation}
- 2 g^2 \, N \, (N^2 - 1) \; \, \mbox{Tr}
\raisebox{-2.6mm}{\scriptsize{$\!\!\!\!\!\!\!\!\!\!$soft} P} \;\,
\frac{P \cdot K}{P^2} \, I_{\, 1 \, 1}^{+ +} \, .
\label{2gh}
\end{equation}
Comparing (\ref{2gh}) with the \lq\lq $d_6$ term'' in
(\ref{22d}) and noting that $K$ ($P$) is hard (soft), we see
that (\ref{2gh}) is subleading.

Figs. 2(j) and (k): It can easily be shown that the sum of the
contributions of Fig. 2(j) and Fig. 2(k) vanishes.

Figs. 3(a) and (b): The contribution is
\[
12 g^2 \, N_F \, (N^2 - 1) \; \mbox{Tr}
\raisebox{-2.6mm}{\scriptsize{$\!\!\!\!\!\!\!\!\!\!$soft} P}
\sum_{\sigma = \pm} \displaystyle{\raisebox{0.6ex}{\scriptsize{*}}}
\! \tilde{S}^{(\sigma)} (P) \, (k_0 - \sigma {\bf k} \cdot
\hat{{\bf p}}) \, I_{\, 1 \, 1}^{+ + } \, .
\]
This is of the same type as (\ref{q-1}) with the second term in the
square brackets and is nonleading as compared to (\ref{q044}).

This completes the proof of (\ref{Pressure2}).
\section{Summary and discussion}

Applying the HTL resummation scheme to the one-loop correction to
the composite vertex $\langle P | \Theta_{\mu \nu} | P \rangle$, we
have deduced the energy density and the pressure density, in the
high-$T$ limit, for soft gluons and soft quarks in a hot quark-gluon
plasma. The resultant energy density for soft gluons,
${\cal E}_g (p)$, and for soft quarks, ${\cal E}_q (p)$, are given
in (\ref{g-final1}) and (\ref{q044}), respectively. In contrast to
the soft-gluon case, to leading order at high-$T$, the energy
density for soft quarks, ${\cal E}_q (p)$, does not receive
contributions from HTL's. This difference originates in different
statistics. To leading order, the hydrostatic pressure density
${\cal P} (p)$ is related to the energy density ${\cal E} (p)$
through ${\cal P} (p) \simeq {\cal E} (p) / 3$, for both soft-gluon
and soft-quark sectors. Thus the relation that is characteristic of
free-massless-particle systems still holds approximately.

Now, as in Sec. IV, we discuss the physical content of $\langle
\Theta_{0 0} \rangle$ by decomposing it into the sum of matrix
elements of $S \otimes S^*$. We take up (\ref{g0-1}) and
(\ref{fig2a1}), and expands
$\displaystyle{\raisebox{0.6ex}{\scriptsize{*}}} \!
\Delta_{\alpha \beta} (P)$, Eq. (\ref{g-eff-pro}), in powers of
$\Pi_\ell (P)$:
\begin{eqnarray}
\displaystyle{\raisebox{0.6ex}{\scriptsize{*}}} \!
\Delta_{\alpha \beta} (P) & = &
\displaystyle{\raisebox{0.6ex}{\scriptsize{*}}} \!
\Delta_{\alpha \beta}^{(F)} (P) + \sum_{n = 0}^\infty
\displaystyle{\raisebox{0.6ex}{\scriptsize{*}}} \!
\Delta_{\alpha \beta}^{(n)} (P) \, ,
\label{tennkai0} \\
\displaystyle{\raisebox{0.6ex}{\scriptsize{*}}} \!
\Delta_{\alpha \beta}^{(F)} (P) & \equiv & P_\alpha P_\beta \left\{
\Delta^+ (P) \right\}^2 \, , \nonumber \\
\displaystyle{\raisebox{0.6ex}{\scriptsize{*}}} \!
\Delta_{\alpha \beta}^{(n)} (P) & \equiv & \sum_{\ell = T \, , L}
{\cal Q}_{\alpha \beta}^{(\ell)} \left\{ \Delta^+ (P)
\right\}^{n + 1} \left\{ \Pi_\ell (P) \right\}^n \nonumber \, .
\end{eqnarray}
Inserting (\ref{tennkai0}) into (\ref{g0-1}) and (\ref{fig2a1}), we
have, with obvious notations,
\begin{equation}
\langle \Theta_{0 0}^{(g 0)} \rangle
\rule[-3mm]{.14mm}{7mm} \raisebox{-2mm}{\scriptsize{$\;$Fig. 1(a)}}
= \langle \Theta_{0 0}^{(g 0)} \rangle_F + \sum_{n = 0}^\infty
\langle \Theta_{0 0}^{(g 0)} \rangle_n
\label{ill1}
\end{equation}
and
\begin{equation}
\langle \Theta_{0 0} \rangle
\rule[-3mm]{.14mm}{7mm} \raisebox{-2mm}{\scriptsize{$\;$Fig. 2(a)}}
= \langle \Theta_{0 0} \rangle_F + \sum_{n = 0}^\infty \langle
\Theta_{0 0} \rangle_n \, ,
\label{ill2}
\end{equation}
respectively. According to the general arguments in \cite{baier}
(see also \cite{bay}), through analytic continuations to the
Minkowski space, (\ref{ill1}) and (\ref{ill2}) go to the expressions
written in the real-time formulation:
\begin{eqnarray}
& & \langle \Theta_{0 0}^{(g 0)} \rangle_F = - \frac{i}{2}
\left( N^2 - 1 \right) \int \frac{d^{\, 4} P}{(2 \pi)^4} \,
\theta_{0 0}^{\alpha \beta} (P) P_\alpha P_\beta \,
\frac{\partial}{\partial \lambda^2} D_{1 1} (P; \lambda^2)
\rule[-3mm]{.14mm}{7mm} \raisebox{-2mm}{\scriptsize{$\;\lambda =
0$}} \, ,
\label{irasuto1} \\
& & \langle \Theta_{0 0}^{(g 0)} \rangle_n = - \frac{i}{2} \left(
N^2 - 1 \right) \int
\frac{d^{\, 4} P}{(2 \pi)^4} \, \theta_{0 0}^{\alpha \beta} (P)
\sum_{\ell = T \, , L} {\cal Q}^{(\ell)}_{\alpha \beta}
\nonumber \\
& & \mbox{\hspace*{11ex}} \times
\sum_{j = 1}^2 (-)^{j - 1} D_{1 j} (P) \left\{ \left[ \hat{\Pi}_\ell
(P) \, \tau_3 \, D (P) \right]^n \right\}_{j 1} \, ,
\label{irasuto11} \\
& & \langle \Theta_{0 0} \rangle_F = - \frac{g^2}{2} N \left( N^2 -
1 \right) \int \frac{d^{\, 4} P}{(2 \pi)^4}
\int \frac{d^{\, 4} K}{(2 \pi)^4} \, \theta_{0 0}^{\alpha \beta} (K)
{\cal V}_{\beta \gamma \epsilon} (K, P) \, {\cal V}^\epsilon_{\;\,
\delta \alpha} (K + P, - P) \nonumber \\
& & \mbox{\hspace*{11ex}} \times \sum_{i, \, j = 1}^2 D_{i 1} (K)
D_{1 j} (K) D_{j i} (K + P) P_\gamma P_\delta \,
\frac{\partial}{\partial \lambda^2} D_{i j} (P; \lambda^2)
\rule[-3mm]{.14mm}{7mm} \raisebox{-2mm}{\scriptsize{$\;\lambda =
0$}} \, , \nonumber \\
\\
& & \langle \Theta_{0 0} \rangle_n = - \frac{g^2}{2} N \left( N^2 -
1 \right) \int \frac{d^{\, 4} P}{(2 \pi)^4} \int
\frac{d^{\, 4} K}{(2 \pi)^4} \, \theta_{0 0}^{\alpha \beta} (K)
{\cal V}_{\beta \gamma \epsilon} (K, P) \, {\cal V}^\epsilon_{\;\,
\delta \alpha} (K + P, - P) \nonumber \\
& & \mbox{\hspace*{11ex}} \times \sum_{i, \, j = 1}^2 D_{i 1} (K)
D_{1 j} (K) D_{j i} (K + P) \sum_{\ell = T \, , L}
{\cal Q}^{(\ell)}_{\gamma \delta} \nonumber \\
& & \mbox{\hspace*{11ex}} \times \sum_{j' = 1}^2 (-)^{j' - 1}
D_{i j'} (P) \left\{ \left[ \hat{\Pi}_\ell (P) \, \tau_3 \, D (P)
\right]^n \right\}_{j' j} \, .
\label{irasuto2}
\end{eqnarray}

\noindent Here, as in (\ref{real-time}), $P$ and $K$ are the
four-vectors in the Minkowski space, $D$ is the scalar $2 \times 2$
matrix propagator, $\hat{\Pi}_\ell (P)$ is the scalar $2 \times 2$
vacuum-polarization matrix, and $\tau_3$ is the third Pauli matrix.

Applying the general rules \cite{nie} to each term in
(\ref{irasuto1}) - (\ref{irasuto2}), we can identify the
corresponding set of matrix elements of $S \otimes S^*$. As an
example, we take up $\langle \Theta_{0 0}^{(g 0)} \rangle_3$ in
(\ref{irasuto11}) and $\langle \Theta_{0 0} \rangle_3$ in
(\ref{irasuto2}), and find the corresponding matrix elements.
Thermal amplitudes representing $\langle \Theta_{0 0}^{(g 0)}
\rangle_3$ and $\langle \Theta_{0 0} \rangle_3$ are depicted,
respectively, in Figs. 6(a) and (b). In Fig. 6, $P$ is soft, $K$'s
are hard, and $\lq\lq 1$'', $\lq\lq i_1$'' $-$ $\lq\lq i_6$'' are
the thermal indices. We consider the case where $i_1 = i_2 = i_3 =
i_6 = 1$, $i_4 = i_5 = 2$, and all $k_0$'s are positive. The rules
\cite{nie} tell us that the thermal amplitude, Fig. 6(a) [(b)],
under consideration represents $S \otimes S^*$ as depicted in Fig.
7(a) [(b)].

In $S \otimes S^*$ in Fig. 7(a), the \lq\lq probe'' $\Theta_{0 0}$
is put on the soft line $P$ in the $S$-matrix element side, while in
Fig. 7(b), $\Theta_{0 0}$ is on the hard line $K$. From the very
definition (\ref{thermal-ave}), $\langle \Theta_{0 0} \rangle$ is
the (thermal) expectation value of $\Theta_{0 0}$ {\em in the
initial states} --- the states at the left end and the right end of
Fig. 7, which consist of hard active gluons only. Here \lq\lq active
gluons'' means those gluons which directly participate in the
thermal reaction. We recall here that, as has been stressed by
Weldon \cite{tsuika} in a different context, intermediate states or
virtual particles are a mathematical figment of Feynman-Dyson
perturbation theory. Then, Fig. 7(a) [(b)] represents the
contribution {\em through the figment, intermediate soft} [{\em
hard}] {\em gluon with $P$} [{\em $K$}]{\em , to the energy density
of the quark-gluon plasma,} which is composed of interacting gluons
and quarks and is in thermal equilibrium. Thus Fig. 7(a) [(b)] has
nothing to do with {\em the energy of the soft} [{\em hard}] {\em
gluon}. As a matter of course, Fig. 6 or the original Figs. 1, 2,
and 3 include the diagrams, $S \otimes S^*$'s, in which
$\Theta_{0 0}$ is put on the gluon line in the initial state.

In this relation, we may view Figs. 2 and 3 as giving the correction
to the energy-momentum density of hard gluons and quarks. To the
lowest order, the energy density of hard gluons is obtained from
(\ref{g02}) by setting $\rho_T (\xi) = \rho_L (\xi) = \epsilon (\xi)
\, \delta (\xi^2 - p^2)$,
\begin{equation}
{\bf P}_0^{(g 0)} \rule[-3.3mm]{.14mm}{8mm}
\raisebox{-3mm}{\scriptsize{\mbox{\hspace{0.07mm}
\scriptsize{hard gluon}}}} = 2 (N^2 - 1) \int
\frac{d^{\, 3} k}{(2 \pi)^3} \, k \, n_B (k) \, .
\label{hard}
\end{equation}
The contributions from Figs. 2(a) - (f) to the energy density are of
the following two types:
\begin{eqnarray}
& & g^2 T \int \frac{d^{\, 3} p}{(2 \pi)^3} \, \frac{1}{p^2 +
(3 m_g^2)} \int \frac{d^{\, 3} k}{(2 \pi)^3} \, \frac{1}{k} \,
n_B (k) \, ,
\label{A-type} \\
& & g^2 \int \frac{d^{\, 3} p}{(2 \pi)^3} \, \frac{1}{p^2 +
(3 m_g^2)} \int \frac{d^{\, 3} k}{(2 \pi)^3} \, n_B (k) \{ 1 +
n_B (k) \} \, . \nonumber \\
\label{B-type}
\end{eqnarray}
Carrying out the integration over ${\bf k}$ in each contribution, and
summing over the contributions from all other diagrams, we obtain
(\ref{g-final}) and (\ref{g-final1}). In
order to obtain the correction to ${\bf P}^{(g 0)}_0
\rule[-3.3mm]{.14mm}{8mm}
\raisebox{-3mm}{\scriptsize{\mbox{\hspace{0.07mm}
\scriptsize{hard gluon}}}}$, Eq. (\ref{hard}), from Figs.
2(a) - (f), we carry out the integrations over soft-p in
(\ref{A-type}) and (\ref{B-type}):
\begin{eqnarray}
\mbox{Eq. (\ref{A-type})} & \propto & g^2 \, p^* \, T \int
\frac{d^{\, 3} k}{(2 \pi)^3} \, \frac{1}{k} \, n_B (k) \, ,
\label{1A-type} \\
\mbox{Eq. (\ref{B-type})} & \propto & g^2 \, p^* \int
\frac{d^{\, 3} k}{(2 \pi)^3} \, n_B (k) \{ 1 + n_B (k) \} \, .
\label{1B-type}
\end{eqnarray}
Here $p^*$ denotes the boundary between the soft- and hard-$P$
regions, $g T << p^* << T$. Eqs. (\ref{1A-type}) and (\ref{1B-type})
is ${\cal O} (p^* / T)$ smaller than (\ref{hard}) in the hard $k$
$\{ = {\cal O} (T) \}$ region, which gives the leading contributions
to the integrals in (\ref{hard}), (\ref{1A-type}), and
(\ref{1B-type}). Thus Figs. 2(a) - (f) yield ${\cal O} (p^* / T)$
smaller corrections to the hard-gluon energy density. This is also
the case for Figs, 2(g) - (k). The diagrams Figs. 3(a) - (c) yield
higher-order corrections to
${\bf P}_0^{(q 0)} \rule[-3.3mm]{.14mm}{8mm}
\raisebox{-3mm}{\scriptsize{\mbox{\hspace{0.07mm}
\scriptsize{hard quark}}}}$.
\setcounter{equation}{0}
\setcounter{section}{1}
\section*{Appendix A Effective propagators}
\def\theequation{\mbox{\Alph{section}.\arabic{equation}}}

In this Appendix, we display the spectral representations for the
effective propagators.

\hspace*{2ex}

\noindent {\em Effective gluon propagator} (cf. (\ref{g-eff-pro}))

\hspace*{2ex}

\begin{eqnarray}
& & \frac{1}{P^2 - \Pi_{T / L} (P)} \nonumber \\
& & \mbox{\hspace*{3.4ex}} = - \int_0^{1 / T} d \tau \,
e^{p_0 \tau} \int_{- \infty}^{+ \infty} d \xi \, \rho_{T / L} (\xi)
\left( 1 + n_B (\xi) \right) e^{- \xi \tau} \, , \nonumber \\
\label{spect1} \\
& & \rho_{T / L} (\xi) \equiv - \frac{1}{\pi} \, Im \frac{1}{P^2 -
\Pi_{T / L} (P)} \rule[-4.4mm]{.14mm}{10mm}
\raisebox{-3.6mm}{\scriptsize{$\, p_0 = \xi + i 0^+$}} \, ,
\label{spect3}
\end{eqnarray}
where $n_B (\xi)$ is as in (\ref{bose}). Explicit expression of
$\Pi_{T / L} (P)$ is given in \cite{gluon}. The spectral function,
$\rho_{T / L} (\xi)$, defined above satisfy various sum rules, among
which we use the following ones in the text;
\begin{eqnarray}
& & \int_{- \infty}^{+ \infty} d \xi \, \xi^n \, \rho_{T/L} (\xi)
= n \, , \;\;\;\;\;\; (n = 0, 1) \, ,
\label{sum-rule1} \\
& & \int_{- \infty}^{+ \infty} \frac{d \xi}{\xi} \, \rho_\ell (\xi)
= \frac{1}{p^2 + 3 \, \delta_{\ell L} \, m_g^2} \, , \;\;\;\;\;\;
(\ell = T, L) \, .
\label{sum-rule2}
\end{eqnarray}
In (\ref{sum-rule2}), $m_g$ is as in (\ref{gluon-mass}).

\hspace*{2ex}

\noindent {\em Effective quark propagator} (cf. (\ref{S-hat}))

\hspace*{2ex}

\begin{eqnarray}
& & \displaystyle{\raisebox{0.6ex}{\scriptsize{*}}} \!
\tilde{S}^{(\sigma)} (P) = - \int_0^{1 / T} d \tau \, e^{p_0 \tau}
\int_{- \infty}^{+ \infty} d \xi \, e^{- \xi \tau} \rho_\sigma (\xi)
\nonumber \\
& & \mbox{\hspace*{29.5ex}} \times \left( 1 - n_F (\xi) \right)
\;\;\;\; (\sigma = \pm) \, , \nonumber \\
\label{spect2} \\
& & \rho_\sigma (\xi) \equiv - \frac{1}{\pi} \, Im
\displaystyle{\raisebox{0.6ex}{\scriptsize{*}}} \!
\tilde{S}^{(\sigma)} (P) \rule[-4.4mm]{.14mm}{10mm}
\raisebox{-3.6mm}{\scriptsize{$\, p_0 = \xi + i 0^+$}}
\, ,
\label{spect4}
\end{eqnarray}
where $n_F (\xi)$ is as in (\ref{fermi}). Explicit expression of
$\displaystyle{\raisebox{0.6ex}{\scriptsize{*}}} \!
\tilde{S}^{(\sigma)} (P)$ is given in \cite{quark}. The spectral
function, $\rho_\sigma (\xi)$, defined above satisfy the sum
rules,
\begin{eqnarray}
& & \int_{- \infty}^{+ \infty} d \xi \, \rho_\sigma (\xi) =
\frac{1}{2} \, ,
\label{sum-rule3} \\
& & \int_{- \infty}^{+ \infty} d \xi \, \xi \, \rho_\sigma (\xi)
= \frac{1}{2} \, \sigma p \, ,
\label{sum-rule4} \\
& & \int_{- \infty}^{+ \infty} d \xi \, \xi^2 \, \rho_\sigma
(\xi) = \frac{1}{2} \, \left( p^2 + m_f^2 \right) \, ,
\label{sum-rule5} \\
& & \int_{- \infty}^{+ \infty} d \xi \, \xi^3 \, \rho_\sigma
(\xi) = \frac{1}{2} \, \sigma \, p \left( p^2 + \frac{5}{3} \, m_f^2
\right) \, ,
\label{sum-rule6}
\end{eqnarray}
where $m_f$ is as in (\ref{quark-mass}).
\setcounter{equation}{0}
\setcounter{section}{2}
\section*{Appendix B Derivation of (19)}
\def\theequation{\mbox{\Alph{section}.\arabic{equation}}}

In this Appendix, we deduce (\ref{real-time}) from (\ref{rei0}).
Using the \lq\lq $\tau$-representations'' (\ref{tau}) and
(\ref{tau1}) (in Appendix C) for $\Delta^+ (K + P)$ and $\left\{
\Delta^+ (K) \right\}^2$, after standard manipulations, we obtain
\begin{eqnarray}
\langle p | \, \frac{1}{2} \, \phi^2 (0) \, | p \rangle & = & -
\frac{g^2}{2} \; \mbox{Tr}
\raisebox{-2.6mm}{\scriptsize{$\!\!\!\!\!\!\!$} P} \left\{
\Delta^+ (K) \right\}^2 \Delta^+ (K + P) \nonumber \\
& = & \frac{g^2}{16} \int_{- \infty}^{+ \infty}
\frac{d^{\, 3} p}{(2 \pi)^3} \, \frac{1}{E k^2} \sum_{\tau = \pm}
\sum_{\sigma = \pm} \left[ \left\{ (1 + n) \left( \frac{1 +
\sigma}{2} + n' \right) \nonumber \right. \right. \\
& & \left. \left. \mbox{\hspace*{30ex}}
- n \left( \frac{1 - \sigma}{2} + n'\right)
\right\} \right. \nonumber \\
& & \mbox{\hspace*{11ex}} \left. \times \left\{ \frac{1}{k} \,
\frac{1}{k + \sigma E + \tau p_0} + \frac{1}{(k + \sigma E + \tau
p_0)^2 } \right\} \right. \nonumber \\
& & \mbox{\hspace*{11ex}} \left. + \frac{n (1 + n)}{T} \,
\frac{\sigma}{k + \sigma E + \tau p_0} \right] \, ,
\label{rei}
\end{eqnarray}
where $n \equiv n_B (k)$ and $n' \equiv n_B (E)$ with $E =
|{\bf k} + {\bf p}|$. We note that (\ref{rei}) may be written as
\begin{eqnarray}
\mbox{(\ref{rei})} & = & - \frac{g^2}{8} \frac{\partial}{\partial
\lambda^2} \int \frac{d^{\, 3} k}{(2 \pi)^3} \frac{1}{E E_\lambda}
\nonumber \\
& & \times \sum_{\tau = \pm} \sum_{\sigma = \pm} \frac{\left(
1 + n_B (E_\lambda) \right) \left( \frac{1 + \sigma}{2} + n' \right)
- n_B (E_\lambda) \left( \frac{1 - \sigma}{2} + n'
\right)}{E_\lambda + \sigma E + \tau p_0}
\rule[-3.5mm]{.14mm}{11mm} \raisebox{-3mm}{\scriptsize{$\; \lambda =
0$}} \, ,
\label{rei-1}
\end{eqnarray}

\noindent where $E_\lambda \equiv \sqrt{k^2 + \lambda^2}$. In the
form (\ref{rei}) or (\ref{rei-1}), we can continue $p_0 \; (= 2 \pi
i n T)$ to real energy $ p_0 + i 0^+$. Straightforward calculation
shows that
\begin{eqnarray}
\mbox{(\ref{rei-1})} & = & \frac{i}{2} \, g^2 \int
\frac{d^{\, 4} K}{(2 \pi)^4} \sum_{i = 1}^2 (-)^{i - 1} D_{i 1}
(K + P) \, \nonumber \\
& & \mbox{\hspace*{17ex}} \times \frac{\partial D_{1 i} (K ;
\lambda^2)}{\partial \lambda^2}
\rule[-3.5mm]{.14mm}{10mm} \raisebox{-3mm}{\scriptsize{$\; \lambda =
0$}} \, .
\label{rei-2}
\end{eqnarray}
The formula (\ref{rei-2}) is written in terms of real-time thermal
$\phi^3$ theory \cite{lan}, which is formulated on the time path
$- \infty$ $\to$ $+ \infty$ $\to$ $- \infty$ $\to$ $- \infty - i /
T$, in a complex time plane. $D_{1 i} (K ; \lambda^2)$ is the bare
thermal propagator of a boson with mass $\lambda$. In passing, the
result (\ref{rei-2}) is in accord with the general argument of
analytic continuation of thermal amplitudes (cf. \cite{baier,bay}).

Employing the mass derivative formula \cite{nie3,nie-sem} for
$\partial D_{1 i} / \partial \lambda^2$, we obtain
\begin{eqnarray}
\mbox{(\ref{rei-2})} & = & \frac{1}{2} g^2 \int
\frac{d^{\, 4} K}{(2 \pi)^4} \sum_{i, \, j = 1}^2 (-)^{i + j}
D_{1 i} (K) \nonumber \\
& & \mbox{\hspace*{17.3ex}} \times D_{i j} (K) D_{j 1} (K + P) \, .
\label{rei-3}
\end{eqnarray}
\setcounter{equation}{0}
\setcounter{section}{3}
\section*{Appendix C List of useful formulae}
\def\theequation{\mbox{\Alph{section}.\arabic{equation}}}

In this Appendix we collect various formulae used in the text. For
the purpose of deriving those formulae, we introduce the \lq\lq
$\tau$-representations'' \cite{bra1,pis8} for the bare propagators.
As in Sec. I, following \cite{bra1}, we use an index $r$, $r = +$
for bosons and $r = -$ for fermions:
\[
n^r = \left\{
\begin{array}{ll}
n^+ (k) = n_B (k) \, , & \mbox{for bosons} \\
n^- (k) = n_F (k) \, , & \mbox{for fermions}
\end{array}
\right.
\]
and introduce
\begin{eqnarray*}
& & f^r_+ (k) = 1 + r n^r (k) \, , \;\;\;\;\;\; f^r_- = r n^r (k)
\, , \\
& & g^r (k) = r n^r (k) \left( 1 + r n^r (k) \right) \, .
\end{eqnarray*}
Then, the boson and fermion propagators (cf. (\ref{mode})) may be
written as
\begin{eqnarray}
& & \Delta^r (K) = - \frac{1}{2 k} \int_0^{1 / T} d \tau \, e^{k_0
\tau} \sum_{s = \pm} f_s^r (k) e^{- s k \tau} \, ,
\label{tau} \\
& & \left\{ \Delta^r (K) \right\}^2 = - \frac{1}{2 k^2}
\frac{1}{K^2} \nonumber \\
& & \mbox{\hspace*{12.6ex}} + \frac{1}{4 k^2 T} \, g^r (k)
\int_0^{1 / T} d \tau \, e^{k_0 \tau} \left[ e^{- k \tau} +
e^{k \tau} \right] \nonumber \\
& & \mbox{\hspace*{12.6ex}} + \frac{1}{4 k^2} \int_0^{1 / T} \tau \,
d \tau \, e^{k_0 \tau} \sum_{s = \pm} s f_s (k) e^{- s k \tau} \, .
\nonumber \\
\label{tau1}
\end{eqnarray}

\hspace*{2ex}

\noindent {\em Formulae for evaluating Figs. 2(a) - (f)}

\hspace*{2ex}

As an illustration, we sketch how to evaluate, under the HTL
approximation,
\begin{eqnarray*}
T \sum_{n = - \infty}^{+ \infty} \frac{k^2}{P^2 - \Pi_\ell (P)}
\otimes I^{+ +}_{\, 1 \, 1} & = & T \sum_{n = - \infty}^{+ \infty}
\frac{1}{P^2 - \Pi_\ell (P)} \; \mbox{Tr}
\raisebox{-2.6mm}{\scriptsize{$\!\!\!\!\!\!\!$} K} \; k^2 \Delta^+
(K) \Delta^+ (K + P) \, , \nonumber \\
& & \mbox{\hspace*{34ex}} (\ell = T, \, L) \, .
\end{eqnarray*}

\noindent Using (\ref{tau}) for $\Delta^+$'s, standard manipulations
lead to
\begin{eqnarray}
& & \mbox{Tr} \raisebox{-2.6mm}{\scriptsize{$\!\!\!\!\!\!\!$} K}
\; k^2 \Delta^+ (K) \Delta^+ (K + P) \nonumber \\
& & \mbox{\hspace*{4ex}} = \frac{1}{4} \int \frac{d^{\, 3}
k}{(2 \pi)^3} \frac{k}{E} \sum_{\tau = \pm} \left[ \frac{1 +
n_B (k) + n_B (E)}{k + E + \tau p_0} \right. \nonumber \\
& & \mbox{\hspace*{25.8ex}} \left. + \frac{n_B (E) - n_B (k)}{k -
E + \tau p_0} \right] \, ,
\label{ex1}
\end{eqnarray}
where $E = |{\bf k} + {\bf p}|$. We take out from (\ref{ex1}) the
UV-divergent piece;
\begin{equation}
\frac{1}{4} \int \frac{d^{\, 3} k}{(2 \pi)^3} \, \frac{k}{E}
\sum_{\tau = \pm} \frac{1}{k + E + \tau p_0} \, .
\label{ex2}
\end{equation}
Continuing the form (\ref{ex2}) to real $p_0$, we can easily show
that
\[
\mbox{(\ref{ex2})} = - i \int_{- \infty}^{+ \infty}
\frac{d k_0}{2 \pi} \int \frac{d^{\, 3} k}{(2 \pi)^3} \frac{k^2}{K^2}
\frac{1}{(K + P)^2} \, ,
\]
which is nothing but the \lq\lq $T = 0$ sector'' of (\ref{ex1}).
Thus (\ref{ex1}) is not the HTL and we ignore it.

Using (\ref{spect1}) for $(P^2 - \Pi_\ell (P))^{- 1}$, under the HTL
approximation, we have
\begin{eqnarray}
& & T \sum_{n = - \infty}^{+ \infty} \frac{k^2}{P^2 - \Pi_\ell (P)}
\otimes I^{+ +}_{1 1} \nonumber \\
& & \mbox{\hspace*{4ex}} \simeq - \frac{1}{4}
\int_{- \infty}^{+ \infty} d \xi \, \rho_\ell (\xi) \int
\frac{d^{\, 3} k}{(2 \pi)^3} \sum_{\tau = \pm} \left[ \frac{1}{k}
\, n_B (k) \, n_B (\xi) \right. \nonumber \\
& & \mbox{\hspace*{7ex}} \left. - \tau \frac{(1 + n_B (k)) n_B (k)
- \tau \, (n_B (k) - n_B (E)) n_B (\xi)}{{\bf p} \cdot \hat{{\bf k}}
- \tau \xi} \right] \, .
\label{ex4}
\end{eqnarray}

\noindent Since $\xi, \; p << k$, we can employ the approximations,
\begin{eqnarray}
& & n_B (k) - n_B (E) \simeq n_B (k) (1 + n_B (k)) \frac{{\bf p}
\cdot \hat{{\bf k}}}{T} \, , \nonumber \\
& & n_B (\xi) \simeq \frac{T}{\xi} \nonumber \, ,
\end{eqnarray}
to get
\begin{eqnarray}
\mbox{(\ref{ex4})} & = & - \frac{1}{2} \int_{- \infty}^{+ \infty}
\frac{d \xi}{\xi} \, \rho_i (\xi) \int \frac{d^{\, 3} k}{(2 \pi)^3}
\nonumber \\
& & \times \left[ \frac{T}{k} \, n_B (k) + n_B (k) ( 1 + n_B (k))
\right] \nonumber \, .
\end{eqnarray}
Using the spectral sum rule (\ref{sum-rule2}), we finally obtain
\begin{eqnarray*}
T \sum_{n = - \infty}^{+ \infty} \frac{k^2}{P^2 - \Pi_\ell (P)}
\otimes I_{1 1} & \simeq & - \frac{1}{8} \, \frac{T^3}{p^2 + 3 \,
\delta_{\ell L} \, m_g^2} \, , \\
& & \mbox{\hspace*{10ex}} (\ell = T, L) \, ,
\end{eqnarray*}
where $m_g$ is as in (\ref{gluon-mass}).

Evaluation of leading-order contributions of other formulae used in
the text are obtained in a similar manner:
\begin{eqnarray}
& & T \sum_{n = - \infty}^{+ \infty} \frac{1}{P^2 - \Pi_\ell (P)} \,
I^{+}_{\, 1 \, 0} \simeq \frac{1}{12} \, \frac{T^3}{p^2 + 3 \,
\delta_{\ell L} \, m_g^2} \, , \nonumber \\
& &  \mbox{\hspace*{39ex}} (\ell = T, L) \, ,
\label{koushiki1} \\
& & T \sum_{n = - \infty}^{+ \infty} \frac{k^2}{P^2 - \Pi_\ell (P)}
\otimes I^{+}_{\, 2 \, 0} \nonumber \\
& & \mbox{\hspace*{4ex}} \simeq T \sum_{n = - \infty}^{+ \infty}
\frac{k^2}{P^2 - \Pi_\ell (P)} \otimes I^{+ +}_{\, 1 \, 1} \simeq -
\frac{1}{8} \, \frac{T^3}{p^2 + 3 \, \delta_{\ell L} \, m_g^2} \, ,
\nonumber \\
\\
& & T \sum_{n = - \infty}^{+ \infty} \frac{k^4}{P^2 - \Pi_\ell (P)}
\otimes I^{+ +}_{\, 2 \, 1} \simeq \frac{5}{32} \, \frac{T^3}{p^2 +
3 \, \delta_{\ell L} \, m_g^2} \, ,
\\
& & T \sum_{n = - \infty}^{+ \infty} \frac{1}{P^2 - \Pi_L (P)}
\frac{p^2}{P^2} \, I^{+}_{\, 1 \, 0} \simeq - \frac{1}{12} \,
\frac{T^3}{p^2 + 3 \, m_g^2} \, ,
\label{koushiki22} \\
& & T \sum_{n = - \infty}^{+ \infty} \frac{k^2}{P^2 - \Pi_L (P)}
\frac{p^2}{P^2} \otimes I^{+}_{\, 2 \, 0} \nonumber \\
& & \mbox{\hspace*{4ex}} \simeq T \sum_{n = - \infty}^{+ \infty}
\frac{p^2} {P^2 - \Pi_L (P)} \frac{k^2}{P^2} \otimes I^{+ +}_{\, 1
\, 1} \simeq \frac{1}{8} \, \frac{T^3}{p^2 + 3 \, m_g^2} \, ,
\nonumber \\
& & \\
& & T \sum_{n = - \infty}^{+ \infty} \frac{1}{P^2 - \Pi_L (P)}
\frac{(P \cdot K)^2}{P^2} \otimes I^{+ +}_{\, 1 \, 1} \simeq
\frac{1}{24} \, \frac{T^3}{p^2 + 3 \, m_g^2} \, , \nonumber \\
& & \\
& & T \sum_{n = - \infty}^{+ \infty} \frac{1}{P^2 - \Pi_L (P)}
\frac{(P \cdot K)^2}{P^2} \otimes I^{+}_{\, 2 \, 0} \simeq
\frac{1}{24} \, \frac{T^3}{p^2 + 3 \, m_g^2} \, , \nonumber \\
& & \\
& & T \sum_{n = - \infty}^{+ \infty} \frac{{\bf k} \cdot {\bf p}}
{P^2 - \Pi_L (P)} \frac{P \cdot K}{P^2} \otimes I^{+ +}_{\, 1 \, 1}
\simeq - \frac{1}{24} \, \frac{T^3}{p^2 + 3 \, m_g^2} \, , \nonumber
\\
& & \\
& & T \sum_{n = - \infty}^{+ \infty} \frac{{\bf k} \cdot {\bf p}}
{P^2 - \Pi_L (P)} \frac{P \cdot K}{P^2} \otimes I^{+}_{\, 2 \, 0}
\simeq - \frac{1}{24} \, \frac{T^3}{p^2 + 3 \, m_g^2} \, ,
\nonumber \\
& & \\
& & T \sum_{n = - \infty}^{+ \infty} \frac{k^2} {P^2 - \Pi_L (P)}
\frac{(P \cdot K)^2}{P^2} \otimes I^{+ +}_{\, 2 \, 1} \simeq -
\frac{5}{96} \, \frac{T^3}{p^2 + 3 \, m_g^2} \, . \nonumber \\
& & \label{koushiki11}
\end{eqnarray}
In deriving (\ref{koushiki22}) - (\ref{koushiki11}), we have used
(\ref{tau}) for $1 / P^2 = \Delta^+ (P)$.

Using the above formulae, we obtain
\begin{eqnarray}
& & T \sum_{n = - \infty}^{+ \infty} \frac{\left( K |{\cal Q}^{(T)}|
K \right)}{P^2 - \Pi_T (P)} \otimes I^{+ +}_{\, 1 \, 1} \nonumber \\
& & \mbox{\hspace*{4ex}} \simeq T \sum_{n = - \infty}^{+ \infty}
\frac{ \left(K |{\cal Q}^{(T)}| K \right)}{P^2 - \Pi_T (P)} \otimes
I^{+}_{\, 2 \, 0} \simeq \frac{1}{12} \frac{T^3}{p^2} \, , \\
& & T \sum_{n = - \infty}^{+ \infty} \frac{k^2 \left( K | {\cal
Q}^{(T)}| K \right)}{P^2 - \Pi_T (P)} \otimes I^{+ +}_{\, 2 \, 1}
\simeq - \frac{5}{48} \, \frac{T^3}{p^2} \, , \\
& & T \sum_{n = - \infty}^{+ \infty} \frac{\left( K | {\cal Q}^{(L)}
| K \right)}{P^2 - \Pi_L (P)} \otimes I^{+ +}_{\, 1 \, 1} \nonumber
\\
& & \mbox{\hspace*{3ex}} \simeq T \sum_{n = - \infty}^{+ \infty}
\frac{k_0 \, K^\rho \, {\cal Q}_{0 \rho}^{(L)}}{P^2 - \Pi_L (P)}
\otimes I^{+ +}_{\, 1 \, 1} \simeq - \frac{1}{24} \,
\frac{T^3}{p^2 + 3 \, m_g^2} \, , \nonumber \\
\\
& & T \sum_{n = - \infty}^{+ \infty} \frac{\left( K | {\cal Q}^{(L)}
| K \right)}{P^2 - \Pi_L (P)} \otimes I^{+}_{\, 2 \, 0} \nonumber \\
& & \mbox{\hspace*{3ex}} \simeq T \sum_{n = - \infty}^{+ \infty}
\frac{k_0 \, K^\rho \, {\cal Q}_{0 \rho}^{(L)}}{P^2 - \Pi_L (P)}
\otimes I^+_{\, 2 \, 0} \simeq - \frac{1}{24} \, \frac{T^3}{p^2 + 3
\, m_g^2} \nonumber \\
\\
& & T \sum_{n = - \infty}^{+ \infty} \frac{ k^2 \left( K |
{\cal Q}^{(L)} | K \right)}{P^2 - \Pi_L (P)} \otimes
I^{+ +}_{\, 2 \, 1} \simeq \frac{1}{32} \, \frac{T^3}{p^2 + 3 \,
m_g^2} \, , \\
& & T \sum_{n = - \infty}^{+ \infty} {\cal Q}^{(L)}_{i i} \otimes
I^{+}_{\, 1 \, 0} \simeq 0 \, .
\label{koushiki2}
\end{eqnarray}
Here, as in (\ref{q}), $\left( K | {\cal Q}^{(T / L)} | K \right)
\equiv K^\rho \, K^\sigma \, {\cal Q}_{\rho \sigma}^{(T / L)}$.

For evaluating the formula of the type (\ref{bonnrei22}), further
manipulations are not necessary. In fact, with the help of the
trivial identity,
\begin{eqnarray}
& & T \sum_{n = - \infty}^{+ \infty} \frac{1}{P^{2 \ell}} \, G (P,
K) \otimes I^{+ +}_{\, i \, j} \nonumber \\
& & \mbox{\hspace*{2ex}} = \lim_{m_g \to 0} T \sum_{n = -
\infty}^{+ \infty} \frac{1}{P^2 - \Pi_L (P)} \, \frac{1}{P^{2 (\ell
- 1)}} \nonumber \\
& & \mbox{\hspace*{16.4ex}} \times G (P, K) \otimes
I^{+ +}_{\, i \, j} \, , \,\;\; (\ell = 0, 1) \, ,
\label{mg-to-0}
\end{eqnarray}
we can evaluate the l.h.s. of (\ref{mg-to-0}) by using
(\ref{koushiki1}) - (\ref{koushiki2}).

\hspace*{2ex}

\noindent {\em Formulae for evaluating Figs. 2(j) and (k)}

\hspace*{2ex}

When $I_{\, i \, j}^{- -}$ is substituted for $I^{+ +}_{\, i \, j}$
in each of the formulae above, (\ref{koushiki1}) -
(\ref{koushiki2}), we obtain the r.h.s. of that formula multiplied
by the factor $- 1 / 2$. Namely, writing the l.h.s. of
(\ref{koushiki1}) - (\ref{koushiki2}) in a generic form $T
\sum_{- \infty}^{+ \infty} H (P, K) \otimes I^{+ +}_{\, i \, j}$,
we have
\begin{equation}
T \sum_{n = - \infty}^{+ \infty} H (P, K) \otimes
I^{- -}_{\, i \, j} \simeq - \frac{1}{2} \, T \sum_{n = -
\infty}^{+ \infty} H (P, K) \otimes I^{+ +}_{\, i \, j} \, .
\label{fermion}
\end{equation}

\newpage
\begin{description}
\item{FIG. 1.} Tadpole diagrams. The symbol $\otimes$ indicates the
point where the operator $\Theta_{\mu \nu}$ is inserted. The blobs
indicate the (HTL-resummed) effective propagators.
\item{FIG. 2.} Formally ${\cal O} (g^2)$ diagrams that contribute
to the soft-gluon sector. $K$ is hard ($\sim T$) and $P$ is soft
($\leq {\cal O} (g T)$).
\item{FIG. 3.} Formally ${\cal O} (g^2)$ diagrams that contribute
to the soft-quark sector. $K$ is hard ($\sim T$) and $P$ is soft
($\leq {\cal O} (g T)$).
\item{FIG. 4.} One-loop correction to the composite vertex $\langle
P | \phi^2 / 2 | P \rangle$.
\item{FIG. 5.} A contribution to $S \otimes S^*$, which is included
in the thermal Green function Fig. 4. The left-side part of the
final-state cut line (dot-dashed line) is the $S$-matrix element in
{\em vacuum} theory and the right-side part is the complex conjugate
of the $S$-matrix element, $S^*$. A group of particles on top of the
figure stands for the constituents of the heat bath.
\item{FIG. 6.} Diagrams representing the thermal amplitude $\langle
\Theta_{0 0}^{(g0)} \rangle_3$, Eq. (\ref{irasuto11}), and $\langle
\Theta_{0 0} \rangle_3$, Eq. (\ref{irasuto2}). For visual clarity,
we have used solid lines for gluons. Thick lines represent hard
gluons, while thin lines represents soft gluons.
\item{FIG. 7.} Matrix elements of $S \otimes S^*$, which are
included in the thermal diagrams Fig. 6.
\end{description}
\newpage
\begin{center}
Values of the coefficients in (26) - (28) for various diagrams in
Fig. 2.
\end{center}
\begin{tabular}{|cccccccccc|} \hline
Fig. 2
& $c_1^{(T)}$
& $c_2^{(T)}$
& $c_3^{(T)}$
& $c_4^{(T)}$
& $c_5^{(T)}$
& $c_6^{(T)}$
& $c_7^{(T)}$
& $c_8^{(T)}$
& $c_9^{(T)}$ \\ \hline
(a) &2 &-1 &1 &-1 &2 &0 &4 &-8 &-10
\\ \hline
(b) &-2 &2 &0 &0 &0 &-1 &2 &4 &0 \\ \hline
(c) &1 &-2 &0 &0 &0 &0 &0 &0 &0 \\ \hline
(d) &0 &-1 &-1/2 &1 &2 &0 &0 &0 &0 \\ \hline
(e) &0 &0 &0 &0 &0 &- 1 &0 &0 &2 \\ \hline
(f) &0 &0 &0 &0 &0 &1  &-1 &0 &0 \\ \hline
\end{tabular}

\vspace*{1cm}

\noindent \begin{tabular}{|ccccccc|} \hline
Fig. 2
& $d_1$
& $d_2$
& $d_3$
& $d_4$
& $d_5$
& $d_6$ \\ \hline
(a) &0 &0 &0 &3/2 &-3 &-5/2
\\ \hline
(b) &2 &0 &0 &-3/2 &0 &2 \\ \hline
(c) &-2 &0 &0 &1/2 &0 &0 \\ \hline
(d) &-1 &1 &-1/2 &0 &1 &0 \\ \hline
(e) &1/2 &0 &0 &-1/2 &0 &1/2 \\ \hline
(f) &-1  &0 &0 &1/2  &0 &0 \\ \hline
\end{tabular}

\newpage
\begin{thebibliography}{99}
\bibitem{pis1} R. D. Pisarski, Phys. Rev. Lett. {\bf 63}, 1129
(1989).
\bibitem{bra1} E. Braaten and R. D. Pisarski, Nucl. Phys.
{\bf B337}, 569 (1990).
\bibitem{bra2} E. Braaten and R. D. Pisarski, Nucl. Phys.
{\bf B339}, 310 (1990).
\bibitem{fre1} J. Frenkel and J. C. Taylor, Nucl. Phys. {\bf B334},
199 (1990).
\bibitem{app} See, e.g., {\em Banff/CAP workshop on thermal field
theory}, edited by F. C. Khanna, R. Kobes, G. Kunstatter, and H.
Umezawa, (World Scientific, Singapore, 1994).
\bibitem{wel0} H. A. Weldon Can J. Phys. {\bf 71}, 300 (1993).
\bibitem{fre-bla} F. T. Brandt, J. Frenkel, and J. C. Taylor, Nucl.
Phys. {\bf B410}, 3 (1993).
\bibitem{nai} V. P. Nair, Phys. Rev. D {\bf 48}, 3432 (1993).
\bibitem{blaizot} J.-P. Blaizot and E. Iancu, Nucl. Phys.
{\bf B421}, 565 (1994).
\bibitem{eff-act1} J. C. Taylor and  S. M. H. Wong, Nucl. Phys.
{\bf B346}, 115 (1990); E. Braaten, in {\em Hot Summer Daze,} edited
by A. Gocksch and R. D. Pisarski (World Scientific, Singapore 1992),
p.14; E. Braaten and R. D. Pisarski, Phys. Rev D {\bf 45}, R1827
(1992);
R. Efraty and V. P Nair, Phys. Rev. Lett. {\bf 68}, 2891 (1992).
See also, J.-P. Blaizot and E. Iancu, Phys. Rev. Lett. {\bf 70},
3376 (1993); Nucl. Phys. {\bf B417}, 608 (1994).
\bibitem{eff-act2} F. Frenkel and J. C. Taylor, Nucl. Phys.
{\bf B374}, 156 (1992);
\bibitem{lan} N. P. Landsman and Ch. G. van Weert, Phys. Rep. {\bf
145}, 141 (1987).
\bibitem{gluon} V. P. Silin, Zh. Eksp. Teor. Fiz. {\bf 38}, 1577
(1960) [Sov. Phys. JETP {\bf 11}, 1136 (1960)]; O. K. Kalashnikov
and V. V. Klimov, Yad. Fiz. {\bf 31}, 1357 (1980) [Sov. J. Nucl.
Phys. {\bf 31}, 699 (1980)]; V. V. Klimov, Zh. Eksp. Teor. Fiz.
{\bf 82}, 336 (1982) [Sov. Phys. JETP {\bf 55}, 199 (1982)]; H. A.
Weldon, Phys. Rev D {\bf 26}, 1394 (1982), R. D. Pisarski, Physica A
{\bf 158}, 146 (1989).
\bibitem{quark} V. V. Klimov, Sov. J. Nucl. Phys. {\bf 33}, 934
(1981); H. A. Weldon, Phys. Rev. D {\bf 26}, 2789 (1982);
{\it ibid}. D {\bf 40}, 2410 (1989).
\bibitem{kel} P. F. Kelly, Q. Liu, C. Lucchesi, and C. Manuel,
Phys. Rev. Lett. {\bf 72}, 3461 (1994); Phys. Rev. D {\bf 50}, 4209
(1994). See also, J.-P. Blaizot and E. Iancu, Nucl Phys. {\bf B434},
662 (1995).
\bibitem{nie-sem} A. J. Niemi and G. W. Semenoff, Ann. Phys. (N.Y.),
{\bf 152}, 105 (1984).
\bibitem{nie} A. Ni\'{e}gawa, Phys. Lett. B {\bf 247}, 351 (1990);
N. Ashida, H. Nak\-kagawa, A. Ni\'{e}gawa, and H. Yokota, Phys. Rev.
D {\bf 45}, 2066 (1992); Ann. Phys. (N.Y.) {\bf 215}, 315 (1992) [E:
{\bf 230}, 161 (1994)];
Asida N., Int. J. Mod. Phys. {\bf A 8}, 1729 (1993).
\bibitem{baier} R. Baier and A. Ni\'egawa, Phys. Rev. D {\bf 49},
4107 (1994).
\bibitem{bay} G. Baym and N. D. Mermin, J. Math. Phys. {\bf 2}, 232
(1961); R. Kobes, Phys. Rev. D {\bf 42}, 562 (1990); T. S. Evans,
Nucl. Phys. {\bf B374}, 340 (1992); P. Aurenche and T. Becherrawy,
{\it ibid}. {\bf B379}, 259 (1992); J. C. Taylor, Phys. Rev. D
{\bf 47}, 725 (1992).
\bibitem{tsuika} H. A. Weldon, Phys. Rev. D {\bf 45}, 352 (1992).
\bibitem{nie3} A. Ni\'{e}gawa, Phys. Rev. D {\bf 40}, 1199 (1989).
\bibitem{pis8} R. D. Pisarski, Nucl. Phys. {\bf B309}, 476 (1988).
\end{thebibliography}
\end{document}